\documentclass[aps, prd, 10pt, twocolumn, superscriptaddress,noshowpacs, preprintnumbers, 
nofootinbib,bibnotes,floatfix]{revtex4-2}
\pdfoutput=1

\usepackage[dvipsnames]{xcolor}
\usepackage[colorlinks=true,breaklinks=true]{hyperref}
\hypersetup{allcolors=[rgb]{0.0 0.0 0.70},linkcolor=[rgb]{0.75 0.05 0.05}}
\usepackage{orcidlink}
\usepackage{microtype}
\usepackage{amsmath}
\usepackage{amsfonts}
\usepackage{amssymb}
\usepackage{mathtools}
\usepackage{graphics}
\usepackage{tabularx}
\usepackage[export]{adjustbox}
\usepackage{multirow}
\usepackage{citesort}
\usepackage{graphicx}
\usepackage{url}
\usepackage{soul}
\usepackage{physics}
\usepackage{bm}
\usepackage[dvipsnames]{xcolor}
\usepackage[utf8]{inputenc}
\usepackage{threeparttable}
\usepackage{makecell}
\usepackage{tabularx}

\begin{document}

\title{Interacting Supernovae: \\ a Radio and X-ray Strategy to Constrain the Structure of the Circumstellar Medium}

\author{Shunke Ai \orcidlink{0000-0002-9165-8312}}
\email{shunke.ai@nbi.ku.dk}
\affiliation{Niels Bohr International Academy and DARK, Niels Bohr Institute, University of Copenhagen, Blegdamsvej 17, 2100, Copenhagen, Denmark}

\author{Irene Tamborra \orcidlink{0000-0001-7449-104X}}%
\email{tamborra@nbi.ku.dk}
\affiliation{Niels Bohr International Academy and DARK, Niels Bohr Institute, University of Copenhagen, Blegdamsvej 17, 2100, Copenhagen, Denmark}%

\author{Leonardo Dinoi \orcidlink{0009-0000-9214-5271}}%
\email{leonardo.dinoi@nbi.ku.dk}%
\affiliation{Niels Bohr International Academy and DARK, Niels Bohr Institute, University of Copenhagen, Blegdamsvej 17, 2100, Copenhagen, Denmark}

\date{\today}

\begin{abstract}
The interaction of supernova (SN) ejecta with the dense circumstellar medium (CSM) converts shock kinetic energy into radiation across multiple wavebands. We investigate the dependence of the X-ray and radio emission on the CSM geometry, considering   spherical, hourglass, and disk shapes for the CSM. We find that the spectral and light-curve properties, both in X-ray and radio,  significantly  differ for spherical and non-spherical CSM structures. For a non-spherical CSM, the radio light curve flattens out near the peak frequency, due to efficient free-free absorption by the unshocked  CSM. Moreover, the early rise of the radio light curve is shallower when the CSM density along the observer line of sight is larger than that in other directions.  
If the CSM density is lower along the observer line of sight, the radio light curve flattens near its peak, and the reverse-shock component is negligible in X-rays. Building on these features, we provide a method to constrain the CSM structure based on the rising part the radio light curve in the proximity of its peak; we show that the decay part of the radio light curve, after its peak, carries insight on whether the CSM density profile is wind-like or not. We further  adopt the X-ray signal to corroborate the information extracted from radio.  We test our strategy on SN 1993j and SN 2023ixf. For both SNe, we find that an asymmetric CSM is in excellent agreement with radio and X-ray observations and provides a viable alternative to  non-wind scenarios  suggested in the literature. Our findings highlight the crucial insight provided by radio and X-ray signals into the mass-loss history of the SN progenitor. 
\end{abstract}
\maketitle

\section{\label{sec:introduction}Introduction}

Massive stars eject strong winds throughout their lives and are often surrounded by a dense circumstellar medium (CSM), see e.g.~Refs.~\citep{2014ARA&A..52..487S, 2017hsn..book..875C,2018SSRv..214...27C} for recent reviews. The CSM is shaped by  steady stellar winds emitted during the  life of the collapsing star as well as  impulsive  mass loss episodes  occurring in the year preceding the explosion in  $\mathcal{O}(10\%)$ of massive stars, cf.~e.g.~Refs.~\cite{2014ApJ...780...21M,2022ApJ...924...15J,2022ApJ...924...55G,2014ApJ...781...42O,2021ApJ...907...99S, 2023ApJ...955L...8H, 2024MNRAS.534..271Q}. 

Following the supernova (SN) explosion, the fast-moving and expanding SN ejecta interact with the CSM, driving a  forward shock (FS) propagating in the CSM and  a reverse shock (RS) crossing  the SN ejecta~\citep{1982ApJ...258..790C,1994ApJ...420..268C}. 
The characterization of  this interaction  is crucial to understand the evolution of the massive star as well as its properties. A growing number of observations suggest that the CSM can exhibit large asymmetries and clumpiness~\cite{1996ApJ...472..257B,2001MNRAS.322..100M,1994MNRAS.268..173C,1995MNRAS.276..530C,2002ApJ...572..350F,1993ApJ...405..337B,2026arXiv260630390C}. 
For example, as it propagates in the CSM, the SN shock can interact with gas clumps, caused by dense equatorial winds from  red supergiants, creating asymmetries in the CSM distribution. Wind interactions from the different evolutionary stages of the SN progenitor can also contribute to shape the CSM.

The FS and RS heat the CSM and ejecta, respectively,  converting kinetic energy into thermal energy and, therefore, radiation. As the shock wave sweeps the surrounding CSM,  X-rays are emitted; this radiation can be partially reprocessed, leading to emission in the  optical and ultraviolet bands~\cite{2017hsn..book..875C}. In addition, electrons and protons may be accelerated to relativistic energies through diffusive shock acceleration. These relativistic electrons undergo synchrotron losses and are primarily responsible for the radio emission~\citep{1982ApJ...259..302C, 1998ApJ...499..810C,2016MNRAS.460...44P}. Accelerated relativistic protons  interact with the cold protons of the CSM  and with photons, producing gamma-rays and high-energy neutrinos~\citep{2011PhRvD..84d3003M, 2014MNRAS.440.2528M, 2022JCAP...08..011S, 2023MNRAS.524.3366P,2023PhRvD.108h3035G, 2023PhRvD.108j3033S,2017MNRAS.470.1881P,2025NatRP...7..285T}. Hence, extended multi-wavelength and multi-messenger emission is to be expected from interacting SNe.
The wide range of CSM properties can be responsible for significant diversity in the multi-messenger emission,  in the optical~\citep{2010MNRAS.407.2305V,2016MNRAS.458.2094D,2016MNRAS.458.1253V},   X-ray and radio~\citep{2017ApJ...835..140M,Murase:2023chr, 2025ApJ...979...16I,2026MNRAS.552g1640S,2026arXiv260612846D,2026arXiv260620836W}, and  the neutrino and gamma-ray signals~\cite{2017MNRAS.470.1881P,2022JCAP...08..011S,2023PhRvD.108j3033S,2011PhRvD..84d3003M,Murase:2023chr}. 

SN 1993j and SN 2023ixf are among  the best observed  interacting SNe with asymmetric CSM. SN 1993j~\cite{1993IAUC.5731....1R} was one of the brightest SNe observed after SN 1987A, originating from a K-type supergiant star and detected at $2.6 \pm 0.4$~Mpc from Earth~\cite{1993Natur.364..600S}. The  strong hydrogen emission lines in the spectra of SN 1993j led to its classification as Type II SN; over time, however, such hydrogen lines faded, leading to the appearance of helium lines, characteristic of a Type Ib SN. As a consequence, this SN was  classified as a Type IIb one (i.e., an intermediate class between Type II and Type Ib)~\cite{1993ApJ...417L..71W,1993Natur.364..600S,1994AJ....107.1453B,1993ApJ...415L.103F} with CSM interaction~\cite{1996ApJ...461..993F, 2000AJ....120.1499M, 2009ApJ...699..388C}. SN 2023ixf~\cite{2023TNSAN.119....1P} was  one of the brightest core-collapse events ever observed, originating from a red supergiant and  detected at $6.85$~Mpc from Earth~\cite{2024ApJ...969..126Y}. SN 2023ixf was classified as being of Type IIL~\cite{2023TNSAN.213....1B}. Its hydrogen lines provided evidence of interaction with an extended CSM~\cite{2025MNRAS.539..633M,2025ApJ...985...51N}. Moreover, polarization measurements of  SN 2023ixf indicate that the CSM is non-spherical, with  the density in the polar regions being   about three times larger than the  equatorial one~\citep{2023ApJ...955L..37V, 2025ApJ...982L..32S, 2026ApJ..1000...18V}. Furthermore, the X-ray and radio light curves of SN 1993j and SN 2023ixf hint that their respective CSM density profiles may also deviate from the canonical wind distribution~\citep{1996ApJ...461..993F,2025ApJ...985...51N}.

In this work, we explore the radio and X-ray emission  from interacting SNe, assuming a CSM with spherical, disk, and hourglass shapes. Our goal is to identify the main observational features  in time and energy that could allow us to  constrain the CSM properties. We find that a detailed reconstruction of the temporal evolution of the radio signal is especially important to this scope.

Our paper is arranged as follows. In Sec.~\ref{sec:geo}, we outline the framework to account for different CSM shapes (spherical, disk, and hourglass) as well as our model of the evolution of the shock radius and velocity. We then introduce the modeling of   the X-ray and radio signals as well as our findings on their spectra and  light curves  in Secs.~\ref{sec:radiation-X} and \ref{sec:radio_radiation}. We present a method  to infer the CSM properties in Sec.~\ref{sec:strategy}. In Sec.~\ref{sec:applications}, we  test our method relying on  SN 1993j and SN 2023ixf. Finally, we discuss our findings  and conclude in Sec.~\ref{sec:conclusion}. 

\section{Supernova model setup}
\label{sec:geo}
In this section, we introduce three different shapes  for  the CSM: spherical, disk, and hourglass. We then outline the shock dynamics, the evolution of the shock velocity, and describe the approach adopted to compute the observed electromagnetic flux.

\subsection{Circumstellar medium geometry}
\begin{figure*}
\centering
\includegraphics[width=1.3\columnwidth]{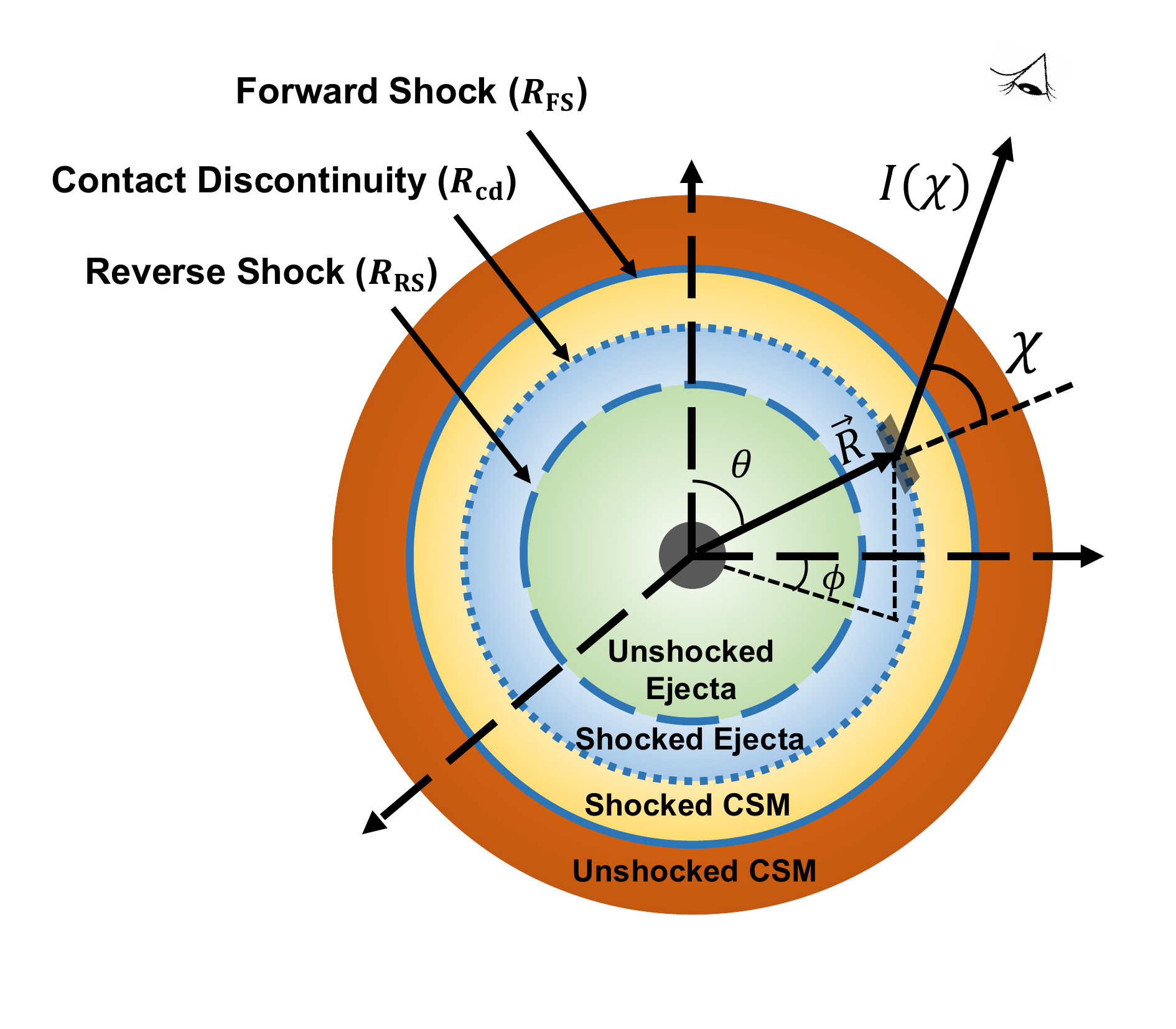} 
\caption{Schematic representation of an interacting SN (not to scale). The central compact object is shown in black, surrounded by the unshocked ejecta (light green), shocked ejecta (light blue), shocked CSM (dark yellow), and unshocked CSM (dark orange). The RS (blue dashed line) is located at the interface between the unshocked and shocked ejecta, while the FS (blue solid line) is located at the interface between the unshocked and shocked CSM. The contact discontinuity (cd, blue dotted line) separates the shocked ejecta and the shocked CSM. Under the thin-shell approximation, the FS, RS, and the contact discontinuity are assumed to have nearly the same radius, i.e., $R_{\rm FS} \approx R_{\rm RS} \approx R_{\rm cd}$ (however, the three radii are well separated in this figure). Each  emitting surface element is characterized by the thin shell radius, $R_{\rm cd}$, that depends on the polar angle $\theta$ defined with respect to $\vec{R}$. The photon intensity from the emitting surface element is $I(\chi)$, where the angle $\chi$ is the relative one between the surface element from which photons are emitted and the  direction of the observer. The latter is assumed to be at a distance $d_L \gg R$ along an arbitrary direction defined by $(\theta_{\rm{obs}}, \phi_{\rm{obs}})$. } 
\label{fig:obs_geo1}
\end{figure*}

Figure~\ref{fig:obs_geo1} provides a schematic representation of an interacting  SN. As the homologously expanding SN ejecta (plotted in light green) propagate in the CSM (in orange), two shocks form, one propagating in the CSM and the other one in the ejecta. 
They are the FS (blue solid line in Fig.~\ref{fig:obs_geo1}) and RS (blue dashed line in Fig.~\ref{fig:obs_geo1}), respectively. The shocked CSM (in dark yellow) and the shocked SN ejecta (in light blue) form a thin shell, whose radial thickness is much smaller than the shock radius. A contact discontinuity (cd, blue dotted line) separates the shocked ejecta and the shocked CSM. Under the thin-shell approximation, we assume that the FS, RS, and the contact discontinuity  have nearly the same radius: $R_{\rm FS}-R_{\rm RS}\ll R_{\rm FS}\approx R_{\rm cd}\approx R_{\rm RS}$, with $R_{\rm cd}$ being the radius of the contact discontinuity. Note, however, that these three radii are well separated  in Fig.~\ref{fig:obs_geo1} to improve its readability. We use $R_{\rm cd}$ to denote the radius of this shell hereafter.

Optical spectropolarimetric observations and the observation of multipeaked hydrogen and helium emission lines for SN 1998s suggest a disk-shaped CSM with a global asphericity exceeding $45\%$~\cite{2000ApJ...536..239L,2000AJ....119.2968G}. A similarly non-spherical  CSM  has  been suggested for SN 2013l~\cite{2017MNRAS.471.4047A} and PTF11iqb~\cite{2015MNRAS.449.1876S}. Interacting SNe with  polar CSM overdensities are relatively rare, but recent spectropolarimetric observations of SN 2023ixf provide hints in this direction~\cite{2025ApJ...982L..32S}. Hence,  we consider three different CSM shapes: spherical, disk, and hourglass.

We model the CSM density profile as
\begin{equation}
\rho_{{\rm CSM}, i}(\theta, R) = D_i(\theta) R^{-s}\, ,
\label{eq:mass_distribution}
\end{equation}
with $\theta$ being the angle between the location of the source emitting surface and  the  polar axis, as shown in  Fig.~\ref{fig:obs_geo1}. To explore the impact of the CSM asymmetry on the electromagnetic emission, we fix $s = 2$, corresponding to a wind-like radial CSM density profile, while varying the CMS angular profile. For the spherical, disk, and hourglass CSM shapes, the $\theta$-dependence of the CSM  density profile is described by the function $D_i(\theta)$:
\begin{eqnarray}
\label{eq:CSMsp}
    D_{\rm sph}(\theta) = D_{\rm sph}(\theta_{\rm obs})\, ,
\end{eqnarray}
\begin{eqnarray}
\label{eq:CSMdisk}
    D_{\rm disk}(\theta) = \frac{D_{\rm disk}(\theta_{\rm obs})}{{\rm exp}\left[-\left\vert\frac{{\pi}/{2}-\theta_{\rm obs}}{\theta_{\rm CSM}}\right\vert^m\right]} {\rm exp}\left[-\left\vert\frac{{\pi}/{2}-\theta}{\theta_{\rm CSM}}\right\vert^m\right]\, ,
\end{eqnarray}
\begin{eqnarray}
\label{eq:CSMhg}
    D_{\rm hg}(\theta) = \frac{D_{\rm hg}(\theta_{\rm obs})}{\left(1+A \left\vert {\rm cos^{\beta}\theta_{\rm obs}} \right\vert\right)} \left(1+A \left\vert {\rm cos^{\beta}\theta} \right\vert\right)\, ;
\end{eqnarray}
here $\theta_{\rm obs}$ denotes the direction of the observer line of sight. The disk shape is inspired by Ref.~\cite{2019ApJ...887..249S}, while the hourglass shape is a  toy model. Unless otherwise specified, we employ the characteristic values listed in Table~\ref{tab:benchmark_value} for the shape parameters in Eqs.~\ref{eq:CSMdisk} and  \ref{eq:CSMhg}. The parameters adopted for the benchmark SN model (assuming Type II SNe) are also presented in this table. 
\begin{table}[t!]
\centering
\caption{Model parameters adopted for our benchmark interacting SN. We fix the ejecta properties according to observations of Type II SNe and the CSM one as from red supergiants observations. }
\renewcommand{\arraystretch}{1.1}
\begin{threeparttable}
\begin{tabular}{lll}
\hline
Physical Quantity & Value & Ref.  \\
\hline
\multicolumn{3}{l}{\textit{Ejecta parameters}} \\
\hline
Kinetic energy ($E_{\rm ej}$) & $10^{51}~{\rm erg}$ &\cite{2003ApJ...582..905H,2022AA...660A..41M} \\
Mass ($M_{\rm ej}$) & $10\, M_{\odot}$ &\cite{2003ApJ...582..905H,2022AA...660A..41M} \\
Outer  density index ($n$) & $12$ & \cite{1999ApJ...510..379M}\\
Inner  density index ($\delta$) & $0.5$ &\tnote{a}   \\
\hline
\multicolumn{3}{l}{\textit{CSM parameters}} \\
\hline
Wind velocity ($v_{\rm w}$) & $25~{\rm km~s^{-1}}$& \cite{2007AA...469..671J}  \\
Radial density index ($s$) & $2$ & \\
Disk  parameter ($m$) & $2$ &\tnote{b}  \\ 
Disk opening angle ($\Theta_{\rm CSM}$) & $60^\circ$ &\tnote{b}  \\
Hourglass  parameter ($A$) & $10$ &\tnote{b} \\
Hourglass  parameter ($\beta$) & $2$ &\tnote{b} \\
\hline
\multicolumn{3}{l}{\textit{Radiation parameters}} \\
\hline
Electron heating efficiency ($\eta_{\rm e, th}$) & $0.5$ & \cite{2025ApJ...993...46W} \\
Electron energy fraction ($\epsilon_e$) & $10^{-2}$ & \cite{2014ApJ...783...91C,2025arXiv250506609M} \tnote{c} \\
Magnetic energy fraction ($\epsilon_B$) & $10^{-3}$ & \cite{2025arXiv250506609M} \tnote{c} \\
Electron power-law index ($p$) & $3$ & \cite{2005ApJ...621..908S,2006ApJ...641.1029C,2010ApJ...725..922S,2019PhRvL.123g1101D} \\
\hline
\end{tabular}
\begin{tablenotes}
\footnotesize
\item[a] Our  results are negligibly affected for $\delta \in [0,1]$.
\item[b] These shape parameters result in a density contrast between the most and least dense regions of  a factor of $10$.
\item[c] The values of $\epsilon_e$ and $\epsilon_B$ may vary according to the environment. We  test the dependence of our results on these two parameters in Sec.~\ref{sec:radio_radiation}.
\end{tablenotes}
\end{threeparttable}
\label{tab:benchmark_value}
\end{table}

For simplicity, we assume azimuthal symmetry for the CSM and define the observer line of sight to have  $\phi_{\rm{obs}} = 0$. Moreover, we select  two representative observer directions, $\theta_{\rm obs} = 0^{\circ}$ and $90^{\circ}$, i.e., configurations such that the observer line of sight aligns with the polar and equatorial directions (cf.~Fig.~\ref{fig:obs_geo1}). 

\begin{figure*}
\includegraphics[width=0.66\columnwidth]{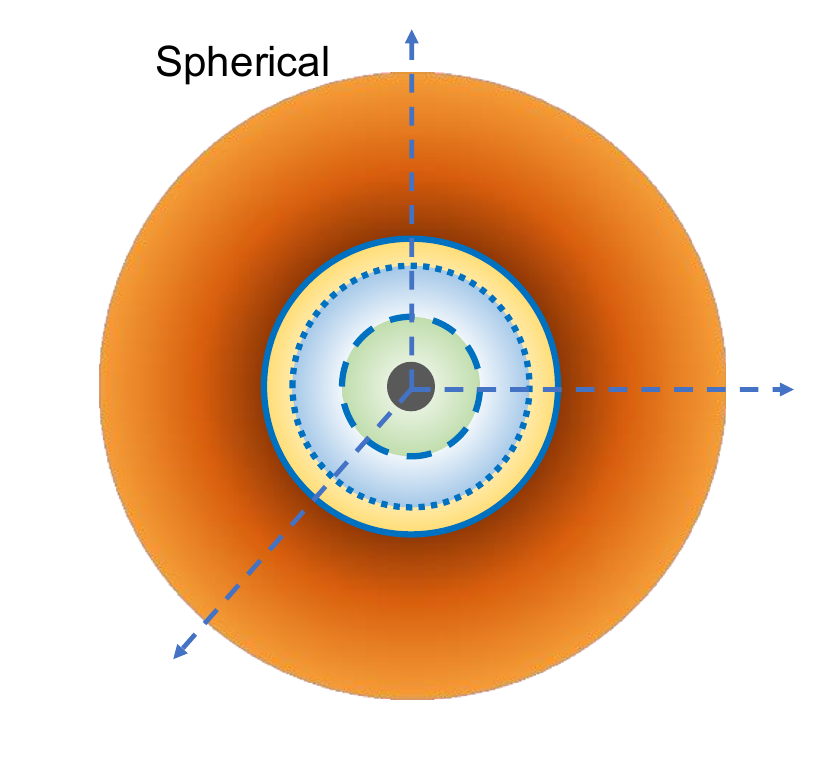} 
\includegraphics[width=0.66\columnwidth]{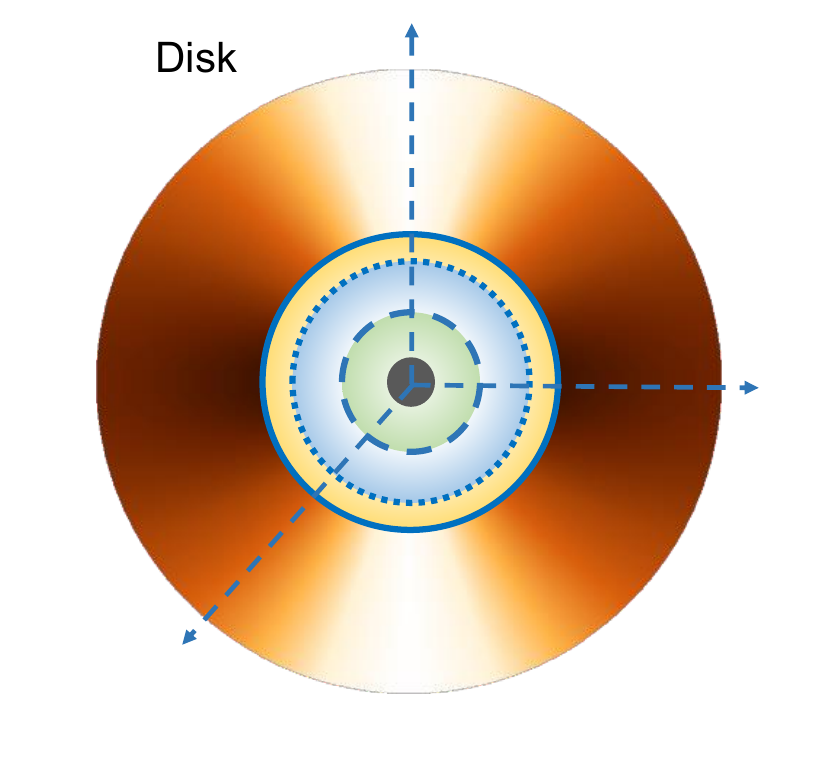} 
\includegraphics[width=0.66\columnwidth]{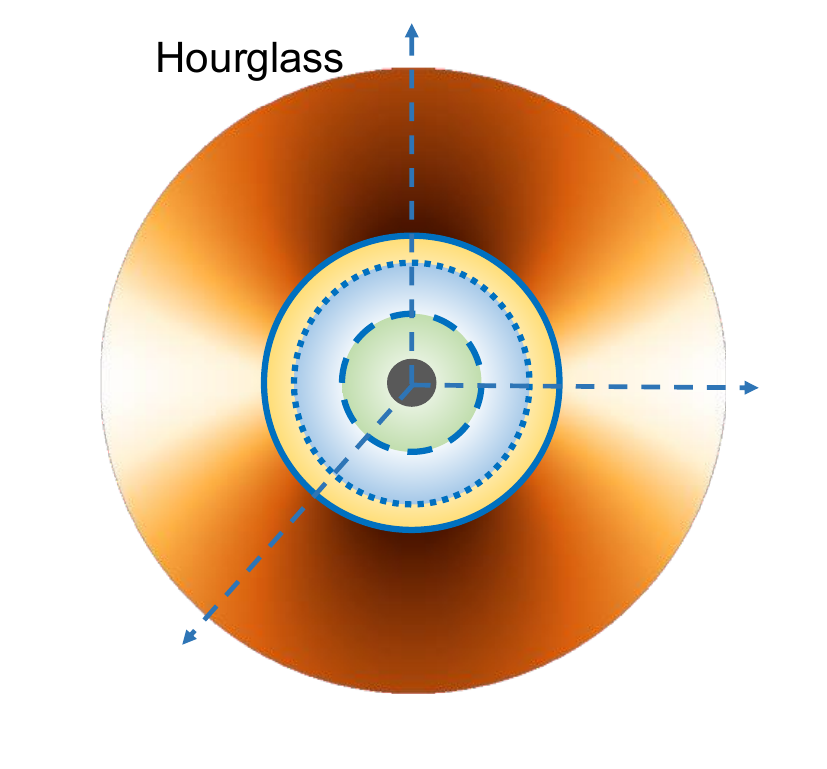}
\caption{Sketch illustrating the spherical, disk and hourglass CSM shapes, from left to right, respectively. The color hues mark the varying CSM density, with darker color indicating regions of higher density. We stress that the disk and hourglass profiles may appear   similar in this 2D sketch, after a $90^\circ$ rotation; however  they are  different in 3D as evident from the functional forms in Eqs.~\ref{eq:CSMdisk} and \ref{eq:CSMhg}.}
\label{fig:obs_geo}
\end{figure*}
Figure~\ref{fig:obs_geo} displays the  spherical, disk, and hourglass  CSM geometries, from left to right, respectively. The spherical CSM geometry is responsible for  uniform emission across different observer directions. On the other hand, the electromagnetic signal observed from a disk- or hourglass-shaped CSM  strongly depends on the observer direction.

\subsection{Shock dynamics}
\label{sec:dynamics}
Under the thin shell approximation, the FS and RS have approximately the same radius. Therefore, the dynamics of the system can be described by the radius ($R_{\rm cd}$) and velocity ($v_{\rm cd}$) of the contact discontinuity between the ejecta and the CSM, see Fig.~\ref{fig:obs_geo1}.
To model the evolution of the radius and velocity for the shocked material, we follow Refs.~\cite{2013MNRAS.435.1520M,1982ApJ...259..302C,1994ApJ...420..268C},
extending their findings to the case of anisotropic CSM.
The mass distribution of the CSM follows Eq.~\ref{eq:mass_distribution}. 
For each angle $\theta$, we apply the  momentum conservation equation  to determine the shock dynamics:
\begin{eqnarray}
\left[\frac{dM_{\rm sh}}{d\Omega}\right]\frac{dv_{\rm cd}}{dt}
&=& R_{\rm cd}^2[\rho_{\rm ej}(v_{\rm ej} - v_{\rm cd})^2 - \rho_{\rm CSM}(\theta)v_{\rm cd}^2] \, , \nonumber \\
\label{eq:momentum_conservation}
\end{eqnarray}
with $v_{\rm cd}$ being the velocity of the contact discontinuity, which is also the bulk velocity of the shocked thin shell. The velocity of the unshocked CSM is assumed to be much smaller than $v_{\rm cd}$, $M_{\rm sh}$ is the mass of the shocked material (including both shocked CSM and ejecta), $v_{\rm ej}$ is the ejecta velocity at the RS, and $\rho_{\rm ej}$ is the ejecta density. 
The mass of the shocked material per unit solid angle ($d\Omega =d\cos\theta d\phi$) is:
\begin{eqnarray}
    \frac{dM_{\rm sh}}{d\Omega} = \int_{R_\star}^{R_{\rm cd}}R^2 \rho_{\rm CSM}(\theta) dr  + \int_{v_{\rm ej}t}^{v_{\rm ej,max}t} R^2 \rho_{\rm ej} dr \, ,
\end{eqnarray}
where $R_\star$ is the SN progenitor radius (and inner boundary of the CSM), and $v_{\rm ej,max}$ is the maximum  ejecta velocity~\footnote{These two quantities do not appear in the final solution under the assumptions that $R_{\rm cd} \gg R_\star$ and $v_{\rm ej} \ll v_{\rm ej,max}$.}. We note that the solutions of Eq.~\ref{eq:momentum_conservation} have the same form as Eqs.~7 and 19 of Ref.~\cite{2013MNRAS.435.1520M}.

Defining $t$ as the time after the explosion and  assuming that the isotropic  SN ejecta follow a double power law, with $\rho_{\rm ej} \propto R^{-n}$ and $\rho_{\rm ej} \propto R^{-\delta}$, inside and outside of the shell with velocity $v_t$~\cite{1999ApJ...510..379M, 2013MNRAS.435.1520M}, the ejecta density can be defined as: 
\begin{widetext}
\begin{eqnarray}
\rho_{\text{ej}}(v_{\text{ej}}, t)
&=&
\left\{
\begin{array}{ll}
\displaystyle
\frac{1}{4\pi(n-\delta)}
\frac{\left[2(5-\delta)(n-5)E_{\text{ej}}\right]^{(n-3)/2}}
{\left[(3-\delta)(n-3)M_{\text{ej}}\right]^{(n-5)/2}}
t^{-3} v_{\text{ej}}^{-n}
& \quad \mathrm{for}\ v_{\text{ej}} > v_t\, , \\[2ex]
\displaystyle
\frac{1}{4\pi(n-\delta)}
\frac{\left[2(5-\delta)(n-5)E_{\text{ej}}\right]^{(\delta-3)/2}}
{\left[(3-\delta)(n-3)M_{\text{ej}}\right]^{(\delta-5)/2}}
t^{-3} v_{\text{ej}}^{-\delta}
& \quad \mathrm{for}\ v_{\text{ej}} < v_t\, ,
\end{array}
\right.
\end{eqnarray}
\end{widetext}
where
\begin{eqnarray}
    v_t = \left[\frac{2(5-\delta)(n-5)E_{\rm ej}}{(3-\delta)(n-3)M_{\rm ej}}\right]^{1/2}\ 
\end{eqnarray}
comes from the density continuity equation at the shell interface, 
 $E_{\rm ej}$ and $M_{\rm ej}$ represent the initial kinetic energy and mass of the ejecta, respectively. 

When the RS propagates in the outer part of the ejecta where $v_{\rm ej} > v_t$, the evolution of the radius of the shocked shell is described by
\begin{widetext}
\begin{eqnarray}
R_{\rm cd}(t, \theta)
&=&
\left[
\frac{(3-s)(4-s)}{4\pi D_i(\theta) (n-4)(n-3)(n-\delta)}
\frac{\left[2(5-\delta)(n-5)E_{\text{ej}}\right]^{(n-3)/2}}
{\left[(3-\delta)(n-3)M_{\text{ej}}\right]^{(n-5)/2}}
\right]^{{1}/{(n-s)}}
t^{{(n-3)}/{(n-s)}}\, ,  
\label{eq:R_sh_front}
\end{eqnarray}
\end{widetext}
where $D_i(\theta)$ has been introduced in Eqs.~\ref{eq:CSMsp}--\ref{eq:CSMhg}.
Correspondingly, the velocity of the shocked shell is 
\begin{widetext}
\begin{eqnarray}
    v_{\rm cd}(t,\theta) = \frac{dR_{\rm cd}}{dt} = \frac{n-3}{n-s}\left[\frac{(3-s)(4-s)}{4\pi D_i(\theta)(n-4)(n-3)(n-\delta)}
\frac{[2(5-\delta)(n-5)E_{\rm ej}]^{(n-3)/2}}
{\left[(3-\delta)(n-3)M_{\rm ej}\right]^{(n-5)/2}}
\right]^{{1}/{n-s}} t^{-{(3-s)}/{(n-s)}}\, .
\end{eqnarray}
\end{widetext}

As the RS  propagates through the inner ejecta with $v_{\rm ej} < v_t$, at later times,  the evolution of the shock radius is given by
\begin{eqnarray}
R_{\rm cd}(t, \theta) = \frac{M_{\rm ej}}{4\pi D_i(\theta)}\left(-1+\sqrt{1+8\pi D_i(\theta)\left(\frac{2E_{\rm ej}}{M_{\rm ej}^3}\right)^{1/2}}\right)\, ,\nonumber \\
\end{eqnarray}
and the velocity of shocked material is 
\begin{eqnarray}
    v_{\rm cd}(t,\theta) =  \sqrt{\frac{2E_{\rm ej}}{M_{\rm ej}}}\left(1 + 8\pi D_i(\theta)\sqrt{\frac{2E_{\rm ej}}{M_{\rm ej}^3}}t\right)^{-1/2}\, .    
\end{eqnarray}
Note that all the solutions for $t > t_t$ are  valid for the wind-like CSM with $s = 2$ assumed in this paper. 

The transition between the two solutions occurs at $t=t_t$, when the reverse shock reaches the ejecta shell with the characteristic velocity $v_{\rm ej}=v_t$. Assuming $R_{\rm cd}(t_t,\theta) = v_t t_t$ (cf.~Eq.~\ref{eq:R_sh_front}), we obtain
\begin{widetext}
\begin{eqnarray}
t_t(\theta) &=& \left[
\frac{(3-s)(4-s)}{4\pi D_i(\theta) (n-4)(n-3)(n-\delta)}
\frac{\left[(3-\delta)(n-3) M_{\text{ej}}\right]^{(5-s)/2}}
{\left[2(5-\delta)(n-5) E_{\text{ej}}\right]^{(3-s)/2}}
\right]^{{1}/{(3-s)}}\ .
\end{eqnarray}
\end{widetext}

When $s = 2$, the spherically symmetric equivalent SN progenitor mass-loss rate is given by 
\begin{eqnarray}
\dot{\cal M}(\theta) = 4\pi D_i(\theta)v_w\, , 
\label{eq:Mdot_eff}
\end{eqnarray}
where $v_w$ represents the SN wind velocity. We normalize the CSM density profile relying on the spherically symmetric equivalent mass-loss rate along the observer's line of sight ($\theta_{\rm obs}$). Thus
\begin{eqnarray}
    &&D_i(\theta_{\rm obs}) = \frac{\dot{\cal M}(\theta_{\rm obs})}{4\pi v_w} \nonumber \\ && = 2.0\times 10^{14}\, {\rm g\,cm^{-1}}\, \left(\frac{\dot{\cal M}(\theta_{\rm obs})}{10^{-4}M_{\odot}\,{\rm yr}^{-1}}\right) \left(\frac{v_w}{25\,{\rm km}\,{\rm s}^{-1}}\right)^{-1}\, . \nonumber \\
    \label{eq:D_theta_obs}
\end{eqnarray}

For $s \neq 2$,  one should follow the shock evolution numerically and the mass loss is unsteady. The spherically symmetric equivalent mass-loss rate is therefore defined at the reference radius $r_0 = 10^{15}\,{\rm cm}$ (cf.~also Ref.~\cite{1996ApJ...461..993F}) as
\begin{eqnarray}
\dot{\cal M}(\theta) = 4\pi D_i(\theta)v_w r_0^{-s+2}\, .
\label{eq:Mdot_eff_sneq2}
\end{eqnarray}
Given $\dot{\cal M}_{\rm eff}(\theta_{\rm obs})$, the corresponding $D_i(\theta_{\rm obs})$ can be calculated relying on Eq.~\ref{eq:D_theta_obs}.

\subsection{Observed flux}
Multi-wavelength emission is expected to be produced in the aftermath of the interaction between the shocked CSM and the SN ejecta~\citep{2017hsn..book..875C,2018SSRv..214...27C}. Due to the asymmetric shape of the CSM, both the emitting CSM surface and its electromagnetic flux  depend on the polar angle $\theta$ (cf.~Fig.~\ref{fig:obs_geo1}). For each emitting surface element, we assume that the radiation is   azimuthally symmetric with respect to the surface normal.

In order to compute the radiation intensity reaching a distant observer, we follow Refs.~\cite{1970stat.book.....M,2014PhRvD..90d5032T}. Assuming that  the flux $F$  is emitted from the  surface element at the position $\vec{R}$ (depending on $R_{\rm cd}$, $\theta$ and $\phi$),  and that $\chi$ is the angle between the  radiating surface
element and the observer direction (cf.~Fig.~\ref{fig:obs_geo1}), the photon intensity along the observer line of sight is
\begin{eqnarray}
    I(\vec{R}, \chi) = \frac{F(\vec{R}, \chi)}{2\pi} \left(1 + \frac{3}{2}{\rm cos}\chi\right)\ ,
    \label{eq:intensity}
\end{eqnarray}
which is valid for $\cos\chi > -2/3$ and reproduces the limb-darkening
eﬀect~\cite{1970stat.book.....M}. We note that this angular distribution relies on the Eddington approximation and is therefore not fully accurate, especially for emission from an optically thin slab. Integrating over the hemisphere facing the observer, we obtain 
\begin{eqnarray}
    F_{\rm obs}(t) =
   \frac{1}{d_L^2} \underset{\substack{\mathrm{visible}\\\mathrm{hemisphere}}}{\int} I(\vec{R},\chi) R_{\rm cd}^2(\theta, t) {\rm sin}\theta {\rm cos} \chi d\theta d\phi \ , 
   \label{eq:F_obs}
\end{eqnarray}
where $d_L$ is the SN luminosity distance.

\section{X-ray signal}
\label{sec:radiation-X}
In this section, we outline the modeling of the X-ray signal. X-ray emission from the FS and RS has been explored extensively in the literature, see e.g.~Refs.~\cite{1984A&A...133..264F, 1994ApJ...420..268C, 1996ApJ...461..993F,2004ApJ...605..823B, 2006A&A...449..171N,2012ApJ...747L..17C,2022ApJ...928..122M,2025ApJ...993...46W}. Here,  we do not perform any detailed modeling of the CSM composition and ionization and focus instead on the CSM  geometry. We  investigate the dependence of the  spectral energy distributions and the light curves on the  CSM geometry for our benchmark SN model and variable spherically symmetric equivalent mass-loss rate.

\subsection{Modeling of the signal}
In the downstream regions of  the FS and  RS, the shocked material is heated and emits photons. Since the CSM is approximately at rest  compared to the shocked material ($v_w \ll v_{\rm cd}$), the FS velocity  can be estimated according to the Rankine–Hugoniot jump conditions: $v_{\rm FS} = v_{\rm cd}(\hat{\gamma}+1)/2$, where $\hat{\gamma} = 5/3$ is the adiabatic index. The ejecta in the immediate upstream region of the RS has a bulk velocity $v_{\rm ej} = R_{\rm cd} / t$. The reverse shock velocity is then calculated as $v_{\rm RS} = [(\hat{\gamma}+1)v_{\rm cd} -(\hat{\gamma}-1)v_{\rm ej}]/2$~\cite{1992pavi.book.....S}. The velocity of the shocked material (in the downstream, $v_{\rm cd}$) is defined in  Sec.~\ref{sec:dynamics}. Note that, hereafter, all physical quantities  depend on $\theta$, except for the efficiency parameters. However, we omit such $\theta$ dependence for the sake of brevity, unless necessary. 

Assuming that a constant fraction $\eta_{\rm e,th}$ of the kinetic energy loss of the fluid across the collisionless shock is transferred to thermal electrons, the shock heating rate per solid angle for both shocks is~\cite{2025ApJ...993...46W}: 
\begin{eqnarray}
    \frac{dL_{\rm sh, FS}}{d\Omega} &=& R_{\rm cd}^2 \eta_{\rm e,th} \frac{2}{(\hat{\gamma}+1)^2} \rho_{\rm CSM} v_{\rm FS}^3\ , \\
    \frac{dL_{\rm sh, RS}}{d\Omega} &=& R_{\rm cd}^2 \eta_{\rm e,th} \frac{2}{(\hat{\gamma}+1)^2}\rho_{\rm ej}(v_{\rm ej} - v_{\rm RS})^3\ .
\end{eqnarray}
We assume that both the CSM and the SN ejecta are made of hydrogen (that is ionized after being shocked).  Hence, in the immediate downstream, the electron temperature at the FS and RS is:
\begin{eqnarray}
    T_{\rm e,FS} &=& \eta_{\rm \rm e,th}\frac{\hat{\gamma}-1}{4 k_B} m_p v_{\rm cd}^2 \nonumber \\
    &=& 5.1\times 10^8\, {\rm K} ~\eta_{\rm e,th} \left(\frac{v_{\rm cd}}{5\times 10^3\,{\rm km\,s^{-1}}}\right)^2 \, ,\\
    T_{\rm e,RS}&=& \eta_{\rm \rm e,th}\frac{\hat{\gamma}-1}{4 k_B} m_p (v_{\rm ej} - v_{\rm cd})^2 \nonumber \\
    & = & 6.2 \times 10^6\, {\rm K} ~\eta_{\rm e,th} \left(\frac{v_{\rm cd}}{5\times 10^3\,{\rm km\,s^{-1}}}\right)^2 \, ,
\end{eqnarray}
where $k_B$ denotes the Boltzmann constant, $m_p$ is the proton rest mass, and $T_{e, \rm RS}$ is estimated from  $v_{\rm ej}=[(n-s)/(n-3)]v_{\rm cd}=(10/9) v_{\rm cd}$ (which is valid as the RS  propagates through the outer ejecta, $v_{\rm ej}>v_t$). The protons and electrons are in equipartition with the same temperature, as discussed  later in this section.

The hot electrons can cool through  bremsstrahlung radiation and inverse Compton scattering. The corresponding cooling power densities are~\cite{2025ApJ...993...46W}:
\begin{eqnarray}
\label{eq:Ebrem}
\dot{e}_{\rm brem} &=& \rho \kappa_{\rm es} c \left(\frac{32}{\pi^3}\right)^{1/2} \alpha_e \frac{\rho}{m_p} (m_e c^2 k_B T_e)^{1/2} \bar{g}_{\rm ff} \nonumber \\
& = & 1.6 \times 10^{-23} \left(\frac{T_e}{10^8\,{\rm K}}\right)^{1/2}  ~({\rm erg~cm^3~s^{-1}}) ~n_e n_p\, , \nonumber \\
\end{eqnarray}
\begin{eqnarray}
    \dot{e}_{\rm Comp} &=& \rho \kappa_{\rm es} c \frac{4k_B(T_e-T_{\rm rad})}{m_e c^2} u_{\rm rad} \nonumber \\
    &\approx&4.7\times 10^{-24}\,({\rm erg\,cm^3\,s^{-1}})\, \left(\frac{v_{\rm cd}}{5\times 10^3\,{\rm km\,s^{-1}}}\right)^3 \nonumber \\
    &&\times \left(\frac{T_e}{10^8\,{\rm K}}\right)n_p n_{\rm CSM,p}\, ,
\label{eq:comp_nr}
\end{eqnarray}
where the average Gaunt factor is $\bar{g}_{\rm ff} \simeq 1$ for bremsstrahlung photon energy of $h\nu \approx k_{\rm B}T_{\rm e}$~\cite{1986rpa..book.....R}, $m_e$ is the rest mass of the electron,  $\rho$ denotes the density of the emitting region, $\kappa_{\rm es} = 0.4\,{\rm g^{-1}\,cm^{2}}$ is the electron scattering opacity corresponding to the Thomson cross section in a fully-ionized  CSM made of hydrogen, and $\alpha_e$ is the fine-structure constant. The number densities of protons ($n_p$) and electrons ($n_e$) are assumed to be equal.  The temperature and radiation energy density of the seed photons are $T_{\rm rad}$ and $u_{\rm rad}$, respectively. The temperature of the shocked electrons ($T_e$)  can be  $T_{\rm e,FS}$ or $T_{\rm e,RS}$, depending on which shock is  considered. Here, we assume $T_{\rm rad} \ll T_e$ implying that the seed photons are in the optical band and originate from a cooler region behind the shock \cite{2025ApJ...993...46W, 2025ApJ...985...51N}. If $T_{\rm rad} \sim T_e$, inverse Compton cooling becomes negligible (cf.~Eq.~\ref{eq:comp_nr}). 

As for the inverse Compton cooling rate, since the seed photons may originate from both the FS and RS~\footnote{Note that this treatment is valid as long as the shock power dominates over radioactive decay as the primary power source of the SN electromagnetic emission.}, $u_{\rm rad} = [1/(R_{\rm cd}^2 c)](dL_{\rm sh, FS}/d\Omega + dL_{\rm sh, RS}/d\Omega)$. However, when estimating  $\dot{e}_{\rm Comp}$  (Eq.~\ref{eq:comp_nr}), for convenience, we do not distinguish between FS and RS and adopt the general expression for the shock power per unit solid angle: $dL_{\rm sh}/d\Omega=(1/2)\rho_{\rm CSM}R_{\rm cd}^2v_{\rm cd}^3$~\cite{2013MNRAS.435.1520M}, and 
$u_{\rm rad} = (dL_{\rm sh}/d\Omega)/(R_{\rm cd}^2 c)$. 

The bremsstrahlung and Compton cooling timescales of the downstream material are $t_{\rm brem} = 3 n_e k_B T_e / \dot{e}_{\rm brem}$ and $t_{\rm Comp} = 3 n_e k_B T_e/ \dot{e}_{\rm Comp}$, respectively, with $\dot{e}_{\rm brem}$ and $\dot{e}_{\rm Comp}$ defined as in Eqs.~\ref{eq:Ebrem} and \ref{eq:comp_nr}. The total cooling rate is given by $t_{\rm e,c}^{-1} = t_{\rm brem}^{-1} + t_{\rm Comp}^{-1}$.  

Protons and electrons are assumed to be in equipartition.  The equilibrium timescale required to reach this state can be estimated as~\cite{1986RvMP...58....1S,1992pavi.book.....S}
\begin{eqnarray}
    t_{\rm eq} &=& \left(\frac{m_p}{m_e}\right)t_{\rm Max,e} \nonumber \\
    &=& \left(\frac{m_p}{m_e}\right)\left(\frac{0.290}{\rm ln \Lambda}\right)\left[\frac{m_e^{1/2}(k_B T_e)^{3/2}}{n_e q_e^4}\right]\, ,
\end{eqnarray}
and 
\begin{eqnarray}
    \frac{t_{\rm eq}}{t_{\rm brem}} &=& 6.3 \times 10^{-3} \left(\frac{T_e}{10^8 K}\right)\left(\frac{\rm ln \Lambda}{30}\right)^{-1} \\
    \frac{t_{\rm eq}}{t_{\rm Comp}} &=& 4.6\times 10^{-4} \left(\frac{T_e}{10^8 K}\right)^{3/2} \left(\frac{v_{\rm cd}}{5\times 10^{3}\,{\rm km\,s^{-1}}}\right)^3 \nonumber \\ 
\end{eqnarray}
where $t_{\rm Max,e}$  is the timescale for electrons to acquire a Maxwellian distribution, and $q_e$ is the electron charge. The ``Coulomb logarithm,'' ${\rm ln \Lambda}$, is  assumed to have a fiducial value of $30$~\cite{1986RvMP...58....1S}. It can be seen that the timescale for electrons and protons to reach equilibrium is shorter than the cooling timescales--this  justifies the equipartition assumption.

Due to the shock heating, a hot layer immediately after the shock  forms, where electrons do not cool down efficiently. The optical depth of the hot layer behind the FS or the RS is 
\begin{eqnarray}
    \tau_{\rm hot,FS/RS} \sim n_e \sigma_T \lvert v_{\rm cd}-v_{\rm FS/RS}\rvert t_{\rm e,c}\, ,
\end{eqnarray}
where $\sigma_T$ is the Thomson cross section~\cite{2025ApJ...993...46W}. This optical depth corresponds to a Compton $y$ parameter~\citep{1986rpa..book.....R}
\begin{eqnarray}
    y = \frac{4k_B T_e}{m_e c^2} {\rm max}(\tau_{\rm hot}, \tau_{\rm hot}^2)\, .
\end{eqnarray}

For our benchmark SN (cf.~Table~\ref{tab:benchmark_value}), the Compton $y$ parameter of thermal electrons never exceeds unity. Therefore, we neglect the X-ray emission produced by inverse Compton scattering. Although inverse Compton cooling can be significant, the scattered photons only gain a small amount of energy~\footnote{The resultant emission from inverse Compton scattering between the seed photons and  thermal electrons cannot be ignored when the shock velocity exceeds $10^9\,{\rm km\,s^{-1}}$~\cite{2022ApJ...928..122M,2025ApJ...993...46W}. This is not the case for our  benchmark SN (cf.~Table~\ref{tab:benchmark_value}).}.

The FS and RS can be either adiabatic or radiative, depending on the ratio between the cooling and the dynamical timescales of the shocked material. The shocked ejecta and shocked CSM have the same bulk velocity ($v_{\rm cd}$); hence, their dynamical timescales are given by  $t_{\rm dyn} \approx R_{\rm cd}/v_{\rm cd}$. When $t_{\rm e,c} > t_{\rm dyn}$, the shock is adiabatic. In this case,  cooling is inefficient, the downstream region maintains a relatively high temperature, and the thermal pressure is used to perform $pdV$ work during the expansion of the shocked shell. 
When $t_{\rm e,c} < t_{\rm dyn}$, the shock is radiative. The thermal energy of the plasma can be  converted into photons and diffuse out. Consequently, a cool dense shell forms between the hot layer and the contact discontinuity \cite{1994ApJ...420..268C}. 

In the adiabatic regime, the downstream radiation can be described by a single-temperature bremsstrahlung spectrum, while radiation  follows a cooling bremsstrahlung spectrum in the radiative region. 
In summary, the spectrum is~\cite{1992pavi.book.....S, 2025ApJ...993...46W}:
\begin{eqnarray}
F_{\nu} \propto \Big\{ \begin{array}{ll}
{\rm exp}\left(-x\right) \  \quad \quad & \text{adiabatic} \ , \\
\int_{h\nu/2T}^{\infty} dx \frac{e^{-x} K_0(x)}{x} \  \quad  \quad & \text{radiative}\ , \\
\end{array}
\end{eqnarray}
where $x = h\nu / (m_e c^2)$ is a  dimensionless parameter  and $K_0$ is the zeroth modified Bessel function of  second kind. For each direction $\theta$, the spectrum is normalized to the total bolometric flux  ($F_{\rm bol}$), with $\int_0^{\infty} F_{\nu}d\nu = F_{\rm bol} = (dL_{\rm bol}/d\Omega) / R_{\rm cd}^2 = \eta_{\rm{rad}} (dL_{\rm sh}/d\Omega)/R_{\rm cd}^2$, where the radiation efficiency [$\eta_{\rm ra} = t_{\rm brem}^{-1}/(t_{\rm e,c}^{-1} + t_{\rm dyn}^{-1})$] is introduced to connect the adiabatic and radiative regimes. This spectral shape holds for  both the FS and the RS. 

According to our calculations, both the FS and RS are radiative at early times because of the short dynamical timescale associated with $R_{\rm cd}$, and they transition to the adiabatic regime at later times. For the benchmark parameters  in Table~\ref{tab:benchmark_value} and $\dot{\cal M}(\theta) \sim 10^{-4} M_{\odot}\,{\rm yr}^{-1}$, the transition to the adiabatic regime occurs at about one week for the FS and  one year for the RS. One should expect this transition to occur earlier for a lower CSM density. A cool dense shell can form between the two hot layers. In the radiative regime, the shocked material can cool efficiently, causing the cool dense shell to become thicker than the hot layers.  On the other hand, the shocked ejecta are much more massive than the shocked CSM. Therefore, we approximate the mass of the cool dense shell with the mass of the shocked ejecta. Following the shock dynamics described in Sec.~\ref{sec:dynamics}, the mass of the shocked ejecta per unit solid angle is~\citep{2013MNRAS.435.1520M}:
\begin{widetext}
\begin{eqnarray}
    \frac{dM_{\rm RS}(t)}{d\Omega} = \int_{v_{\rm ej}/t}^{v_{\rm ej,max}/t} r^2\rho_{\rm ej} dr 
    =\left\{\begin{array}{ll}
    \frac{t^{n-3}}{4\pi(n-\delta)(n-3)R_{\rm cd}^{n-3}}\frac{[2(5-\delta)(n-5)E_{\rm ej}]^{(n-3)/2}}{\left[(3-\delta)(n-3)M_{\rm ej}\right]^{(n-5)/2}} \   \quad \quad & t < t_t \, ,  \\
    \frac{M_{\rm ej}}{4\pi} - \frac{R_{\rm cd}^{3-\delta}}{4\pi(n-\delta)(3-\delta)t^{3-\delta}} \frac{[2(5-\delta)(n-5)E_{\rm ej}]^{(\delta-3)/2}}{\left[(3-\delta)(n-3)M_{\rm ej}\right]^{(\delta-5)/2}} \  \quad \quad & t\geq t_t\, . \ 
    \end{array}
    \right.
\end{eqnarray}
\end{widetext}
The mass of the cool dense shell per unit solid angle is approximated as $dM_{\rm CDS}/d\Omega = dM_{\rm RS} / d\Omega$ in the radiative regime of the RS, but is unchanged after the transition to the adiabatic RS regime. The cool dense shell is   made of neutral shocked ejecta~\citep{1994ApJ...420..268C}.

The photons generated at the RS propagate through the cool dense shell, where  bound-free absorption occurs. The bound-free interaction cross section is~\cite{1986rpa..book.....R}:
\begin{eqnarray}
    \sigma_{\rm bf}(\nu) = \left\{ 
    \begin{array}{ll}
    \sigma_{\rm bf,thres}\left(\frac{h\nu}{\varepsilon_{\rm thres}}\right)^{-3} & \quad h\nu \geq \varepsilon_{\rm thres} \, , \\
     0    &  \quad h\nu < \varepsilon_{\rm thres} \, ,
    \end{array}\right.
\end{eqnarray}
where  $\varepsilon_{\rm thres} = 13.6\, {\rm eV}$ is the threshold ionizing energy and $\sigma_{\rm bf, thres} = 6.3 \times 10^{-18}\,{\rm cm}^2$. The corresponding optical depth is $\tau_{\rm bf}(\varepsilon) = \sigma_{\rm bf}(\varepsilon) (dM_{\rm CDS} / d\Omega) / m_p/ R_{\rm cd}^2$. 
The unshocked CSM can also scatter the outward-propagating photons, approximately in the Thomson limit, with an optical depth $\tau_s(\theta) \approx \sigma_TD_i(\theta)/(m_pR_{\rm cd})$, so it will delay the release of photons without  changing the shape of the spectrum. 
Hence, the  flux from the emitting surface  becomes 
\begin{eqnarray}
    F_{\nu}(\vec{R}) = \left\{ \begin{array}{ll}
    F_{\nu}({\vec{R}}) / {\rm max(\tau_s,1)} \  \quad \quad & {\rm FS}\, ,  \\
    F_{\nu}(\vec{R}){\rm exp}[-\tau_{\rm bf}(\nu)] / {\rm max(\tau_s,1)} \  \quad \quad   & {\rm RS}\, .
    \end{array}
    \right. 
\end{eqnarray}
We compute  the observed flux, $F_{\rm \nu,obs}$,  substituting $ F_{\nu}(\vec{R})$  into Eq.~\ref{eq:F_obs}. In the following, to eliminate the dependence on the source distance, we focus on the inferred luminosity,  $L_{\rm \nu}=4\pi d_L^2F_{\rm \nu,obs}$.

\begin{figure}
    \centering
    \includegraphics[width=0.99\columnwidth]{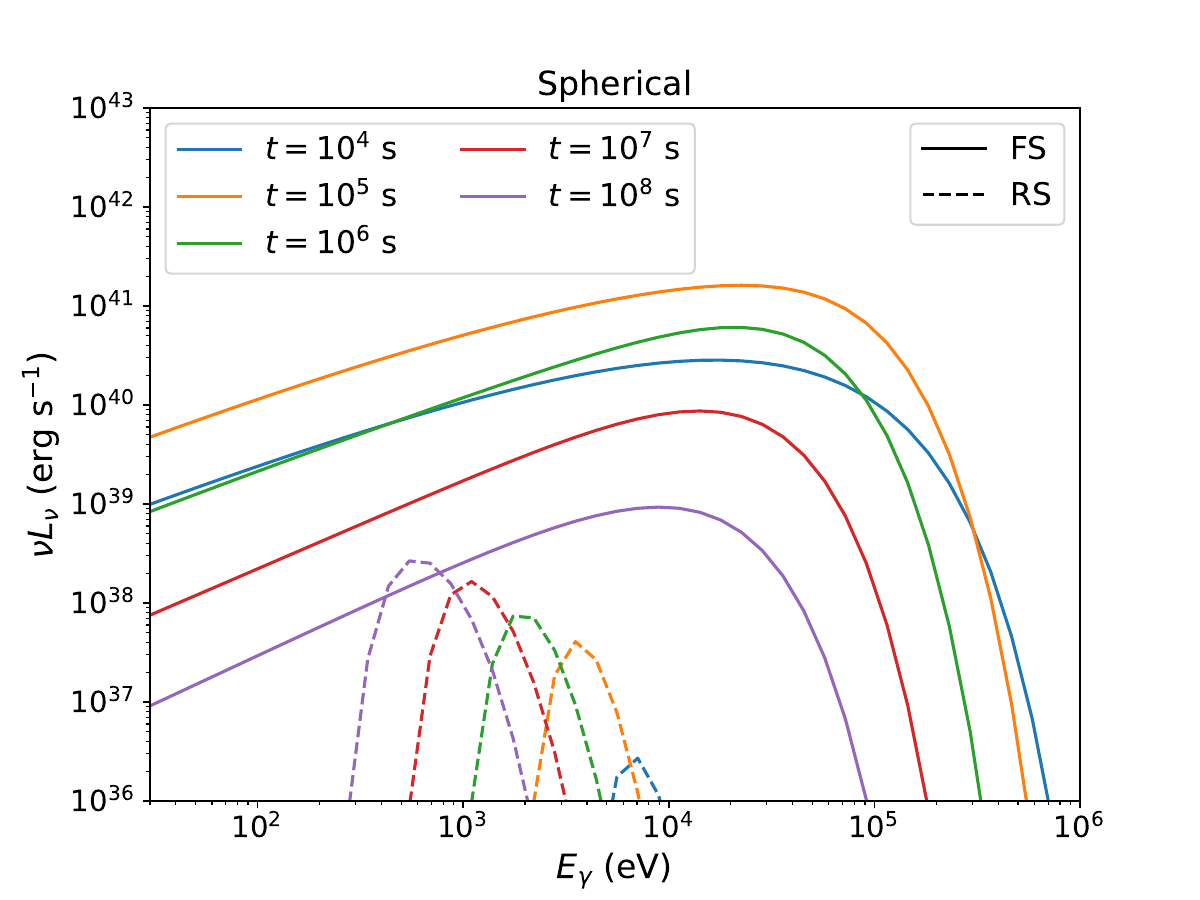} 
    \caption{Spectral evolution of bremsstrahlung radiation, peaking in the X-ray band from  the FS (solid lines) and the RS (dashed lines) from $t=10^4$~s through $10^8$~s after the SN explosion. A spherical CSM with $\dot{\cal M}(\theta_{\rm obs})=10^{-4}\,M_{\odot}$ is adopted; see Table~\ref{tab:benchmark_value}. 
    The FS (RS) emission peaks in the hard (soft) X-ray band. In the  very soft X-ray band, the RS emission  is bound-free absorbed by the cool dense shell.
    }
    \label{fig:Xray_spec}
\end{figure}

Figure~\ref{fig:Xray_spec} shows  the evolution of the  bremsstrahlung spectra from $t=10^4$~s through $10^8$~s for our benchmark SN with a spherical CSM (cf.~Table~\ref{tab:benchmark_value}) and a spherically symmetric equivalent mass-loss rate of $\dot{\cal M}= 10^{-4} M_{\odot}\, {\rm yr}^{-1}$. We distinguish between the contributions  from the RS and the FS.  We can see that the emission from the FS peaks in the hard X-ray band, whereas that from the RS peaks in the soft X-ray band. This difference is due to the  temperatures of the material shocked by the RS and the FS. The RS spectrum  does not extend into the ultraviolet band because of the strong bound-free absorption due to the cool dense shell. At early times, the FS dominates the X-ray emission. As the FS decelerates, both the bolometric luminosity and the temperature of the shocked material decrease. Meanwhile, as the ejecta expand, the density of the cool dense shell decreases. At late times, the RS emission emerges and can  dominate the soft X-ray band.

\subsection{Spectral energy distributions and light curves}
\begin{figure*}
\centering
\includegraphics[width=1.98\columnwidth]{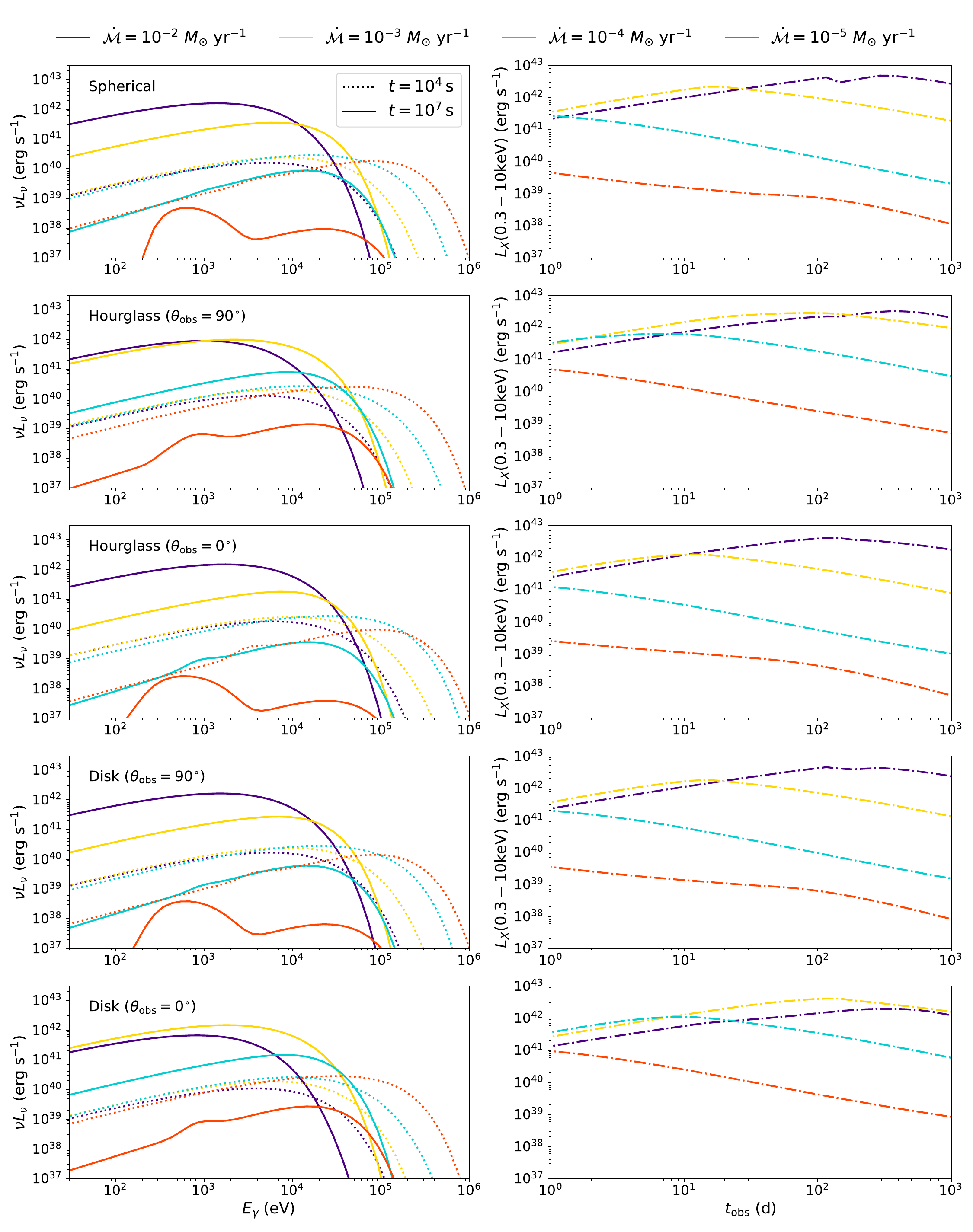}
\caption{X-ray spectra (left) and light curves (right)  for  spherical, hourglass, and disk CSM structures for our benchmark SN, see Table~\ref{tab:benchmark_value}, accounting for the contributions  from the FS and the RS. We consider the following spherically symmetric equivalent mass-loss rates: $\dot{\cal M}(\theta_{\rm obs})=10^{-2}$, $10^{-3}$, $10^{-4}$, and $10^{-5}~M_{\odot}$ (in purple, yellow, cyan, and red, respectively), corresponding to the CSM density profile along the observer line of sight  defined as in Eq.~\ref{eq:Mdot_eff}. The dotted (solid) lines in the left panels correspond to spectra computed at $t=10^4~{\rm s}$  ($10^7~{\rm s}$) after the SN explosion. The  light curves in the right panels are computed in the  $0.3$--$10~{\rm keV}$ energy band. As $\dot{\cal M}(\theta_{\rm obs})$ increases, the peak energy of the X-ray photons decreases, whereas both the peak luminosity and the time of the luminosity peak increase. The effect of the  RS is  only visible at late times  for $\dot{\cal M}(\theta_{\rm obs}) \lesssim 10^{-4}  {M}_\odot\, \rm{yr}^{-1}$.}
\label{fig:Xray_Mdot}
\end{figure*}

Figure~\ref{fig:Xray_Mdot} presents the X-ray spectra and light curves for our three CSM profiles, two selected observer angles, and four spherically symmetric equivalent mass-loss rates along the line of sight ranging from $10^{-5}\, {M}_\odot\, \rm{yr}^{-1}$ to $10^{-2}\, M_\odot\, \rm{yr}^{-1}$. The left panels show the spectra at $t = 10^4\,{\rm s}$ and $10^7\,{\rm s}$. The RS contribution  is not negligible only at late times for a spherically symmetric equivalent mass-loss rate of $\dot{\cal M}(\theta_{\rm obs}) \lesssim 10^{-4} {M}_\odot\, \rm{yr}^{-1}$; in all other cases, the RS emission  can be neglected. The temperature of the FS shocked material decreases with increasing mass-loss rate, resulting in a softer spectral peak. When the CSM density along the line of sight is higher than that in other directions (e.g., for the disk CSM observed from the equator and the hourglass one observed from the pole), the spectral evolution is similar to the one  of the spherical case with the same spherically symmetric equivalent mass-loss rate along the line of sight. When the CSM density along the line of sight is lower than the one along other directions (e.g., for the disk-shaped CSM observed from the pole and the hourglass one observed from the equator), the contribution from other directions dominates. Hence, the spectral energy distribution is similar to the one of the spherical case with a higher spherically symmetric equivalent mass-loss rate. 

The right panels of Fig.~\ref{fig:Xray_Mdot}  represent the light curves in the X-ray band ($0.3$--$10\, {\rm keV}$). A higher spherically symmetric equivalent mass-loss rate results in a larger photospheric radius, below which photons are scattered and their escape is delayed. Therefore, the light curves exhibit a rising phase at early times. In cases without such a rising phase (e.g., $\dot{\cal M}=10^{-5}\, M_{\odot}\, {\rm yr}^{-1}$), a gradual change in the slope of the light curve is still visible, marking the transition from the radiative to the adiabatic shock regimes. Although, at this stage, both the FS and RS are in the adiabatic regime, the luminosity near and after the slope change is dominated by the RS contribution. Cases with   CSM density along the line of sight lower than the one in other directions tend to be brighter than the spherical CSM case for comparable spherically symmetric equivalent mass-loss rates.

\section{Radio signal}
\label{sec:radio_radiation}
In this section, we outline the modeling of the radio signal. We also present our findings on the spectral energy distributions and light curves for the  varying CSM geometries,  spherically symmetric equivalent mass-loss rates, and microphysical parameters.

\subsection{Modeling of the signal}
The radio emission is produced by synchrotron radiation from the electrons accelerated at the FS and propagating into the CSM~\footnote{Note that we only consider CSM-shock synchrotron emission~\cite{2017hsn..book..875C}.}. For a strong non-relativistic shock, the downstream electron number density is $n_{\rm e,d} \approx 4n_{\rm e,u}$, where $n_{\rm e,u} = \rho_{\rm CSM}/m_p$ is the electron number density in the upstream. The downstream internal energy density can be calculated as~\cite{1992pavi.book.....S}:
\begin{eqnarray}
\label{eq:ud}
u_d = \frac{1}{\hat{\gamma}-1}\frac{2}{\hat{\gamma}+1} \rho_{\rm CSM} v_{\rm FS}^2 = \frac{9}{8} \rho_{\rm CSM} v_{\rm FS}^2\ .
\end{eqnarray}
Assuming that a fraction $\epsilon_e$ of the internal energy is dissipated in kinetic energy of electrons due to their random motion and that the accelerated electrons initially follow a power-law energy distribution $dn_{\rm e,acc}/d\gamma\propto \gamma^{-p}$ (see Table~\ref{tab:benchmark_value}), the total number of accelerated electrons in the downstream is 
\begin{eqnarray}
    \frac{dN_{\rm e,acc}}{d\Omega} = \int_{R_{\rm CSM,in}}^{R_{\rm cd}} R^2 dR \int_{\gamma_{\rm min}}^{\gamma_{\rm max}} \frac{1}{4}K_{e,1} \frac{dn_{\rm e,acc}}{d\gamma} d\gamma\, . 
\end{eqnarray}
Here, $R_{\rm CSM,in}$ is the inner radius of the CSM; its value does not affect the final results as long as $R_{\rm cd} \gg R_{\rm CSM,in}$. This is because the number of shocked electrons increases linearly with $R_{\rm cd}$ for $n_{\rm e,CSM} \propto R^{-2}$. The minimum and maximum Lorentz factors of the accelerated electrons are $\gamma_{\rm min}$ and $\gamma_{\rm max}$, respectively. We assume $\gamma_{\rm min}=1$~\footnote{The minimum Lorentz factor is given by $\gamma_{\rm min}=1+{1}/{2}\left({(p-2)}/{(p-1)}\right)\epsilon_e\left({v_{\rm FS}}/{c}\right)^2$~\cite{2022ApJ...932..116H}. This yields $\gamma_{\rm min} \approx 1$ for $p=3$, $\epsilon_e \le 0.1$, and $v_{\rm FS} < 0.1c$, all of which are always satisfied in our model.}. On the other hand, the value of $\gamma_{\rm max}$ negligibly affects the  synchrotron flux, as long as $\gamma_{\rm max} \gg \gamma_{\rm min}$. The normalization coefficient  $K_{e,1}$ is obtained from 
\begin{eqnarray}
    \int_{\gamma_{\rm min}}^{\gamma_{\rm max}}\gamma m_e c^2 K_{e,1}\frac{dn_{\rm e,acc}}{d\gamma}d\gamma = \epsilon_e u_d\, ,
\end{eqnarray}
and for $p > 2$ is
\begin{eqnarray}
K_{e,1} = \frac{\epsilon_e u_d}{m_e c^2} (p-2)\gamma_{\rm min}^{p-2}\, .
\end{eqnarray}

If a fraction $\epsilon_B$ of the internal energy is dissipated into the magnetic field, the magnetic strength in the downstream is
\begin{eqnarray}
    B = \sqrt{8\pi\epsilon_B u_d} \, ,
\end{eqnarray}
with $u_d$ defined as in Eq.~\ref{eq:ud}. The energy distribution of accelerated electrons is affected by the cooling processes, including synchrotron, inverse Compton, and  Coulomb scattering. The  timescales of these processes are~\cite{2017hsn..book..875C,2025ApJ...985...51N}:
\begin{eqnarray}
t_{\rm syn} &=& \frac{6\pi m_e c}
{\sigma_{\rm T} B^2 \gamma} \nonumber \\
& = & 6.2\times 10^{-8}\, {\rm s}~ \rho_{\rm CSM}^{-1} \gamma^{-1} \left(\frac{\epsilon_B}{10^{-3}}\right)^{-1} \nonumber \\ &&\times \left(\frac{v_{\rm cd}}{5\times 10^3\,{\rm km\,s^{-1}}}\right)^{-2}\, , \\
t_{\rm IC} &=& \frac{3 m_e c} 
{4 \sigma_{\rm T} u_{\rm rad}\gamma} \nonumber \\
&=& 1.5 \times 10^{-8}\,{\rm s}~\rho_{\rm CSM}^{-1}\gamma^{-1}\left(\frac{v_{\rm cd}}{5\times 10^3\,{\rm km\,s^{-1}}}\right)^{-3}\,  , \nonumber \\
\\
t_{\rm Coul} &=& 0.67\times 10^{-12}\,{\rm s}~ \gamma \rho_{\rm CSM}^{-1}\, .
\end{eqnarray}

Comparing the  cooling timescales with the dynamical timescale [$t_{\rm dyn}^{-1} = t_{\rm syn}^{-1} + t_{\rm IC}^{-1}$], all electrons with Lorentz factor above the following one are cooled:
\begin{eqnarray}
    \gamma_c &=& \frac{v_{\rm cd}}{R_{\rm cd}}\frac{3m_e c}{\sigma_T}\left[\frac{1}{4R_{\rm cd}^2 c}\left(\frac{dL_{\rm sh}}{d\Omega}\right)+\frac{B^2}{2\pi}\right]^{-1} \nonumber \\
    &\approx& {\rm min}\left(\gamma_{\rm c, syn}, \,\gamma_{\rm c,IC}\right)\, ;
\end{eqnarray}
here we introduced  two characteristic Lorentz factors that can be obtained by setting $t_{\rm dyn}^{-1} = t_{\rm syn}^{-1}(\gamma_c)$ and $t_{\rm dyn}^{-1} = t_{\rm IC}^{-1}(\gamma_c)$:
\begin{eqnarray}
    \gamma_{\rm c,syn} &=& 3.1 \times 10^2 \left(\frac{\rho_{\rm CSM}}{10^{-16}\,{\rm g\,cm^{-3}}}\right)^{-1} \left(\frac{R_{\rm cd}}{10^{15}\,{\rm cm}}\right)^{-1} \nonumber \\
    &&\times \left(\frac{v_{\rm cd}}{5\times 10^3\,{\rm km\,s^{-1}}}\right)^{-1} \left(\frac{\epsilon_B}{10^{-3}}\right)^{-1} \, ,\\
    \gamma_{\rm c,IC} &=& 7.4 \times 10^1 \left(\frac{\rho_{\rm CSM}}{10^{-16}\,{\rm g\,cm^{-3}}}\right)^{-1} \left(\frac{R_{\rm cd}}{10^{15}\,{\rm cm}}\right)^{-1} \nonumber \\
    &&\times \left(\frac{v_{\rm cd}}{5\times 10^3\,{\rm km\,s^{-1}}}\right)^{-2}\, .
\end{eqnarray}
Moreover, assuming  $t^{-1}_{\rm dyn} = t^{-1}_{\rm Coul}$, we have that electrons with $\gamma \lesssim \gamma_{\rm Coul}$ have cooled down, with
\begin{eqnarray}
    \gamma_{\rm Coul} &=&  3.0\times 10^2 \left(\frac{\rho_{\rm CSM}}{10^{-16}\,{\rm g\,cm^{-3}}}\right)\left(\frac{R_{\rm cd}}{10^{15}\,{\rm cm}}\right) \nonumber \\
    &&\times \left(\frac{v_{\rm cd}}{5\times 10^3\,{\rm km\,s^{-1}}}\right)^{-1}\, .
\end{eqnarray}

The energy distribution of accelerated electrons is~\cite{1998ApJ...497L..17S, 2002ApJ...568..820G,2025ApJ...985...51N}: 
\begin{eqnarray}
\frac{dN_{\rm e,acc}/d\Omega}{d\gamma} =  K_{e,2} \left\{ 
\begin{array}{ll}
\gamma_{\rm Coul}^{-1}\gamma^{-(p-1)} &~~~ \gamma_m <\gamma < \gamma_{\rm Coul} \\
\gamma^{-p}& ~~~\gamma_{\rm Coul} < \gamma < \gamma_{c}\, ,\\
\gamma_c \gamma^{-(p+1)} &~~~ \gamma > \gamma_c
\end{array}\right.
\end{eqnarray}
for $\gamma_{\rm min}<\gamma_{\rm Coul}<\gamma_c$, and
\begin{eqnarray}
\frac{dN_{\rm e,acc}/d\Omega}{d\gamma} =  K_{e,2} \left\{ 
\begin{array}{ll}
\gamma_c^{-1}\gamma^{-(p-1)} &~~~ \gamma_m <\gamma < \gamma_c \\
\gamma_c \gamma^{-(p+1)} &~~~ \gamma > \gamma_c
\end{array}\right.
\end{eqnarray}
for $\gamma_{\rm min}<\gamma_c<\gamma_{\rm Coul}$. The normalization coefficient is 
\begin{eqnarray}
    K_{e,2} = \frac{dN_{\rm e,acc}}{d\Omega} \left(\frac{1}{p-1}\frac{1}{\gamma_m^{p-1}}\right)^{-1}\, .
\end{eqnarray}

The local emission flux from  the accelerated electrons is
\begin{eqnarray}
F_{\nu}(R_{\rm cd}) = \frac{1}{R_{\rm cd}^2} \int_{\gamma_{\rm min}}^{\gamma_{\rm max}} P_{\nu,{\rm syn}}(\nu,\gamma)\frac{dN_{\rm e,acc}}{d\Omega} d\gamma\, .
\end{eqnarray}
The synchrotron emission power for a single electron is~\citep{1986rpa..book.....R}
\begin{eqnarray}
P_{\nu,{\rm syn}}(\nu,\gamma) = \frac{\sqrt{3}q_e^3B}{m_e c^2} G\left(\frac{\nu}{\nu_{\rm ch}}\right)\, ,
\end{eqnarray}
where $\nu_{\rm ch} =  {(3\gamma^2 q_e B)}/{(4 \pi m_e c)}$, and $G(x) = x\int_x^{\infty}K_{5/3}(\xi)d\xi$ with $K_{5/3}(\xi)$ being the modified Bessel function of the second kind representing the characteristic frequency of  synchrotron radiation from an electron with a Lorentz factor $\gamma$~\citep{1986rpa..book.....R}. 

In the radio band, synchrotron photons can experience self-absorption in a dense environment. To take this effect into account, we follow Ref.~\cite{2022ApJ...932..116H} (see also Refs.~\cite{1998ApJ...499..810C}). Therefore, the characteristic synchrotron self-absorption frequency is 
\begin{widetext}
\begin{eqnarray}
    \nu_{\rm SSA} = \left\{ \begin{array}{cc}
         R_{\rm cd}^{\frac{2}{(p+5)}} B^{\frac{p+3}{p+5}}t^{-\frac{2}{p+5}} \xi^{-\frac{1}{p+5}} \epsilon^{\frac{2}{p+5}}\eta^{\frac{p+4}{p+5}}\   & \nu_a > \nu_c  \\
        R_{\rm cd}^{\frac{2}{p+4}} B^{\frac{p+6}{p+4}} \epsilon^{\frac{2}{p+4}} \eta \  & \nu_a \le \nu_c
    \end{array} \right.\, ,
\end{eqnarray}
\end{widetext}
where
\begin{eqnarray}
    \xi = \frac{\sigma_T^2}{18\pi m_e c q_e}\, ,
\end{eqnarray}
\begin{eqnarray}
    \eta = \left[\frac{(p-2)\sigma_T}{12\pi^2 m_e^2 c}\right]^{\frac{2}{p+4}}\left(\frac{q_e}{2\pi m_e c}\right)^{\frac{p-2}{p+4}} \, ,
\end{eqnarray}
and $\nu_c$ is the characteristic frequency associated with $\gamma_c$, and $\epsilon = \epsilon_e/\epsilon_B$. Below $\nu_a$, the spectrum is modified as
\begin{eqnarray}
    F_{\nu} = F_{\nu_{\rm SSA}}\left(\frac{\nu}{\nu_{\rm SSA}}\right)^{5/2}\, .
\label{eq:radio_flux_after_ssa}
\end{eqnarray}

As the radio signal propagates in the unshocked CSM, free-free absorption  also needs to be taken into account. Assuming an ionized hydrogen envelope, the free-free absorption  optical depth is 
\begin{eqnarray}
\nonumber
    \tau_{\nu, \rm FFA} &\approx& \int_{R_{\rm cd}}^{\infty}\kappa_{\nu, \rm FFA} n_{\rm e, CSM}^2 dR\, ,
\end{eqnarray}
where the opacity is~\cite{1975A&A....39....1P}
\begin{eqnarray}
    \kappa_{\nu, \rm FFA} = 4.74 \times 10^{-27} \left(\frac{\nu}{1\, {\rm GHz}}\right)^{-2.1} \left(\frac{T_{\rm CSM}}{10^5 {\rm K}}\right)^{-1.35}\, ,
\end{eqnarray}
and $T_{\rm CSM} = 10^5\, {\rm K}$ is assumed~\cite{1996ApJ...461..993F}. For $n_{e,{\rm CSM}}\propto R^{-2}$,  $\tau_{\nu,{\rm FFA}}$ is determined by the local density  near $R = R_{\rm cd}$.
Then, the overall emitted synchrotron flux is 
\begin{eqnarray}
\label{eq:F_FFA}
    F_{\nu, e}({\vec{R}}) = F_{\nu}(\vec{R}){\rm exp}(-\tau_{\rm \nu, FFA}) \, ,
\end{eqnarray}
with $F_{\nu}(\vec{R})$ being the flux after the synchrotron self-absorption correction (Eq. \ref{eq:radio_flux_after_ssa}). The observed flux can be computed by substituting Eq.~\ref{eq:F_FFA} into  Eq.~\ref{eq:F_obs}.

\begin{figure}
    \centering
    \includegraphics[width=0.99\columnwidth]{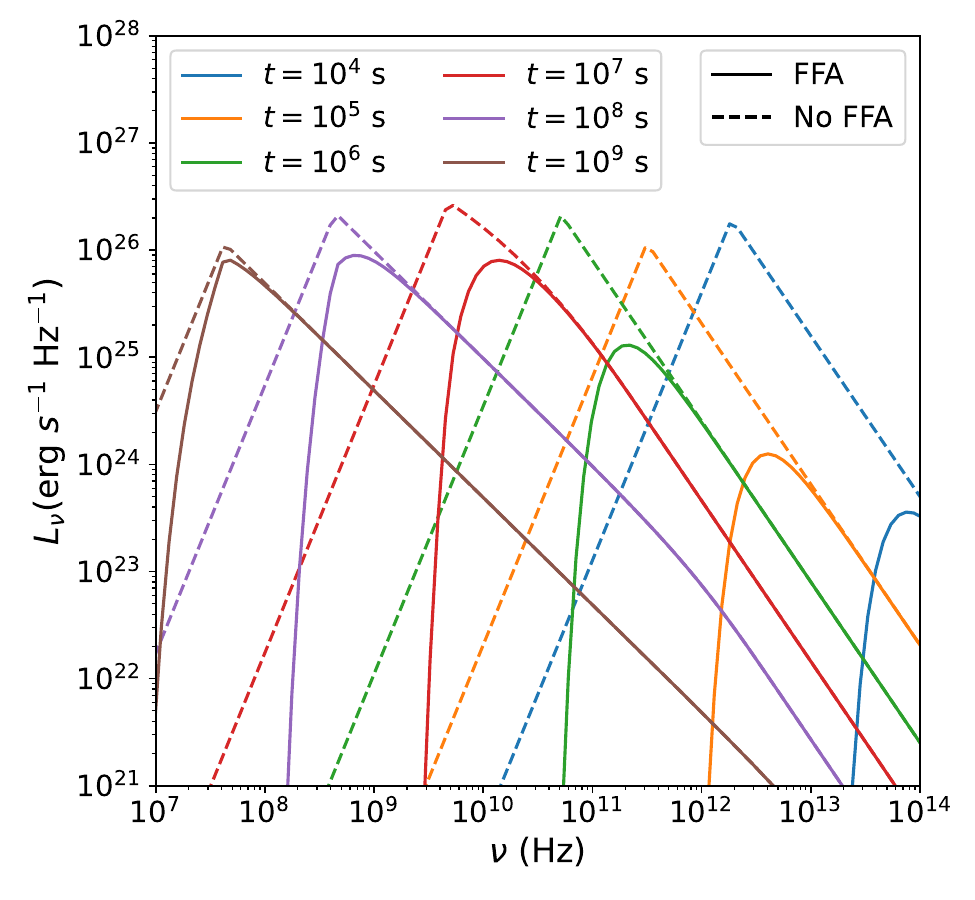} 
    \caption{Spectral evolution of synchrotron radiation from the FS from $10^4$~s through $10^8$~s,  peaking in the radio band. A spherical CSM with $\dot{\cal M}=10^{-4}\, M_{\odot}$ is adopted; all other parameters are set as in Table~\ref{tab:benchmark_value}. Solid and dashed lines represent the spectra with and without free-free absorption (FFA), respectively. Free-free absorption can  modify the synchrotron spectrum significantly.}
    \label{fig:synchrotron_spec}
\end{figure}

Figure~\ref{fig:synchrotron_spec} shows the evolution of  the synchrotron spectrum from $10^4$~s through $10^9$~s for our benchmark SN in a spherical CSM. In order to highlight the free-free absorption  contribution, we plot the spectra with  (without) free-free absorption with solid (dashed) lines. We can see that free-free absorption dominates the entire spectrum at early times,   significantly affecting its shape. As the shock propagates outward, the high-frequency part of the spectrum tends to become transparent. At sufficiently late times, synchrotron self-absorption becomes the dominant absorption mechanism, with  $\nu_a$  decreasing to the meter-wave band. 

\subsection{Spectral energy distributions and light curves}
\begin{figure*}
\centering
\includegraphics[width=1.98\columnwidth]{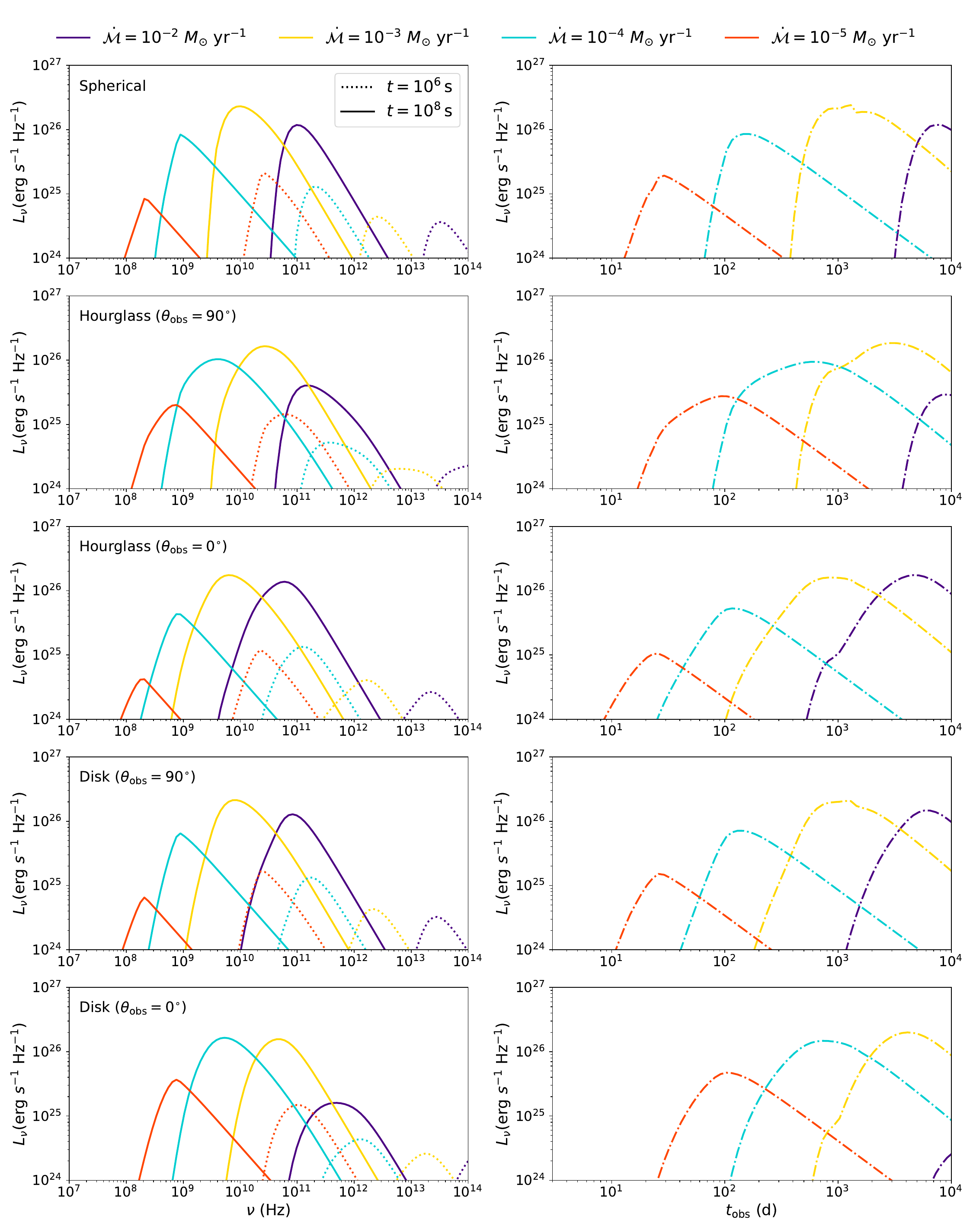}
\caption{Radio spectra (left) and light curves (right) for  spherical, hourglass, and disk CSM geometries for our benchmark SN, see Table~\ref{tab:benchmark_value}, including free-free absorption. We consider the following spherically symmetric equivalent mass-loss rates: $\dot{\cal M}(\theta_{\rm obs})=10^{-2}$, $10^{-3}$, $10^{-4}$, and $10^{-5}~M_{\odot}$ (cf.~Eq.~\ref{eq:Mdot_eff}). The dotted (solid) lines in the left panel represent spectra computed at  $t=10^6~{\rm s}$  ($10^8~{\rm s}$). The light curves in the right panels are computed at  $10~{\rm GHz}$. As $\dot{\cal M}(\theta_{\rm obs})$ increases, the CSM tends to become transparent to radio emission at a later time. Hence,  the peak luminosity is reached later. The shape of the radio light curve is sensitive to the CSM geometry. }
\label{fig:radio_Mdot}
\end{figure*}

Figure~\ref{fig:radio_Mdot} shows the dependence of the spectrum (on the left) and light curve (on the right) on  the spherically symmetric equivalent mass-loss rate for our three CSM geometries and two selected observer directions, from top to bottom, respectively. Note that, due to the strong free-free absorption, we consider $10^6$~s as representative of the early time emission.  Since free-free absorption dominates the spectrum at early times (cf.~Fig.~\ref{fig:synchrotron_spec}), the left panel displays a sharp break in the low-frequency band. However, for $\dot{\cal M}(\theta_{\rm obs}) < 10^{-4}$ and $t \gtrsim 10^8\,{\rm s}$, the low-frequency part of the spectrum slightly deviates from the free-free-dominated regime, and synchrotron self-absorption  becomes significant. 

\begin{figure*}
\centering
\includegraphics[width=1.98\columnwidth]{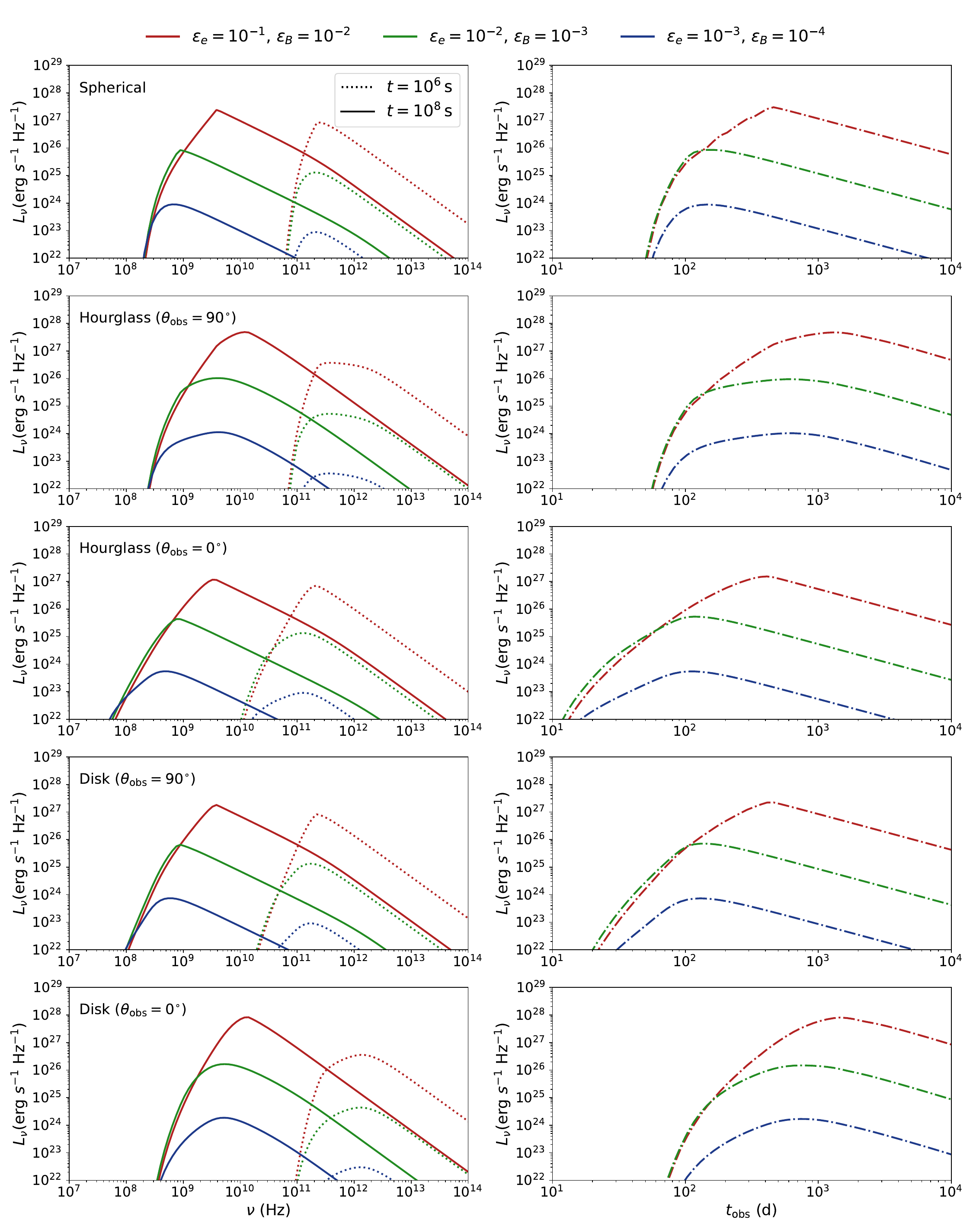}
\caption{Same as Fig.~\ref{fig:radio_Mdot} but  with fixed $\dot{\cal M}(\theta_{\rm obs})=10^{-4}~M_{\odot}~{\rm yr}^{-1}$ and varying microphysical parameters: $(\epsilon_e, \epsilon_B) =(10^{-1},10^{-2})$, $(10^{-2},10^{-3})$, and $(10^{-3},10^{-4})$. As $\epsilon_e$ and $\epsilon_B$ increase, the radio luminosity increases, whereas both the spectral shape and the light curve remain  unchanged.}
\label{fig:radio_eps}
\end{figure*}

The synchrotron spectra are flatten out for asymmetric CSM shapes. This is because the projection in the observer direction averages over the different peak frequencies across the source hemisphere facing the observer. 

The right panels of Fig.~\ref{fig:radio_Mdot} show the radio light curves at $\nu = 10\,{\rm GHz}$. For all CSM geometries and spherically symmetric equivalent mass-loss rates, the rising part of the light curves at early times is dominated by free-free absorption, which   depends on the CSM density. Therefore, the rising slope of the signal is strongly linked to the CSM geometry. 

For a spherical CSM, the light curve rises steeply until its peak. When the CSM density is higher along the observer line of sight than in other directions (e.g., for the hourglass CSM with $\theta_{\rm obs} = 0^{\circ}$ and the disk CSM with $\theta_{\rm obs} = 90^{\circ}$), the rise of the light curve is shallower than the one obtained for the spherical CSM. When the CSM density is lower along the observer line of sight than in other directions (e.g., for the hourglass CSM with $\theta_{\rm obs} = 90^{\circ}$ and the disk CSM with $\theta_{\rm obs} = 0^{\circ}$), the flux rises steeply at early times and becomes flatter near the peak. This characteristic trend observed in the light curves could serve as an indication for distinguishing different CSM geometries and is further explored in Sec.~\ref{sec:strategy}.

The microphysical parameters, $\epsilon_e$ and $\epsilon_B$, are highly uncertain.
 Therefore, Fig.~\ref{fig:radio_eps} explores the dependence of the radio signal on $\epsilon_e$ and $\epsilon_B$. As expected, cases with larger $\epsilon_e$ and $\epsilon_B$ have higher luminosities. For $(\epsilon_e, \epsilon_B) = (10^{-1}, 10^{-2})$, the shapes of the synchrotron spectra and light curves deviate from those shown in Fig.~\ref{fig:radio_Mdot}. This is because, for larger  $\epsilon_e$ and $\epsilon_B$, synchrotron self-absorption  becomes the dominant absorption mechanism, hindering the clear dependence of the signal on the  CSM geometry.

\section{Method to infer the properties of the circumstellar medium}
\label{sec:strategy}
The rising part of the  radio light curves can be used to distinguish among different CSM geometries. In this section, we present a method aiming at inferring the CSM geometry based on the risetime of the radio light curve. We then show how to use the X-ray signal to  corroborate the radio constraints and discuss the dependence of the decaying part of the radio light curve on the CSM profile.

\subsection{The rise time of the radio light curve}
We  focus on the rising part of the light curve that changes with the CSM geometry and the viewing angle. For a given light curve, we  determine the peak time ($t_{\rm peak}$) and  peak luminosity ($L_{\nu,\rm{peak}}$). We then measure the rise time from $1\%$ to $10\%$ of the peak flux ($\Delta t_{0.01-0.1}$) and from $10\%$ of the peak luminosity to the peak ($\Delta t_{0.1-\rm{peak}}$). These parameters, $\Delta t_{0.01-0.1}$ and $\Delta t_{0.1-\rm{peak}}$, are expected to depend on the CSM geometry when free-free absorption is significant. 

\begin{figure}    
  \includegraphics[width=0.99\columnwidth]{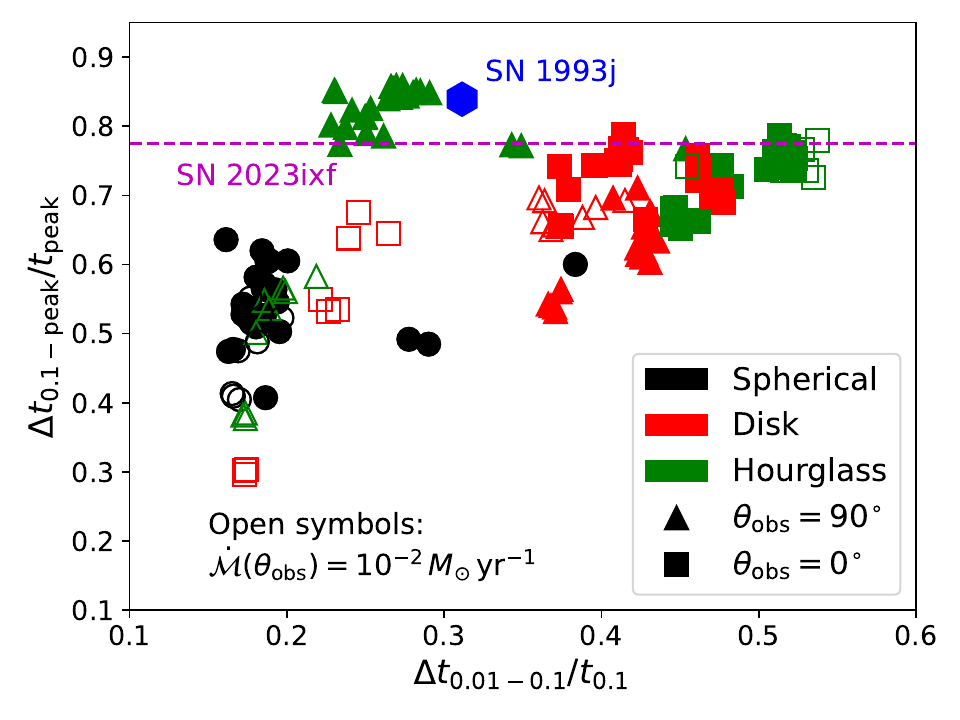} 
   \caption{Scatter plot of the rise times of the radio light curves (at $10$~GHz) for the spherical (black), disk (red), and hourglass (green) CSM geometries. The rise times are computed from $1\%$ to $10\%$ of the peak flux ($\Delta t_{0.01-0.1}$) and from $10\%$ of the peak luminosity to the peak ($\Delta t_{0.1-{\rm peak}}$), normalized by the time at which the flux reaches $10\%$ of the peak and by the peak time, respectively. Triangles and squares represent $\theta_{\rm obs} = 90^{\circ}$ and $\theta_{\rm obs} = 0^{\circ}$, respectively. The light curves are computed for a set of SNe obtained  considering all possible permutations of the following parameters: spherically symmetric equivalent mass-loss rate [$\dot{\cal M}(\theta_{\rm obs}) = 10^{-5}$, $10^{-4}$, $10^{-3}$, and $10^{-2} M_{\odot}~{\rm yr^{-1}}$],  kinetic energy of the ejecta  [$3\times 10^{50}$, $10^{51}~{\rm erg}$, and $3\times10^{51}$], and microphysical parameters [$(\epsilon_e,\epsilon_B) = (10^{-2},10^{-3})$, $(10^{-3},10^{-4})$, and $(10^{-4},10^{-5})$]; all other SN parameters are set as in Table~\ref{tab:benchmark_value}. The rise times cluster according to the CSM shape, except for those with $\dot{\cal M}(\theta_{\rm obs}) =10^{-2}\,M_{\odot}\,\mathrm{yr}^{-1}$, which are marked with open symbols. In addition, we highlight SN~1993j (at $15\,{\rm GHz}$)~\cite{1996ApJ...461..993F} with a blue hexagon and SN~2023ixf (at $10.5\,{\rm GHz}$)~\cite{2025ApJ...985...51N} with a magenta dashed line (because observations do not extend up to  $1\%$ of the peak-luminosity). }
    \label{fig:LC_scatter}
\end{figure}

Figure~\ref{fig:LC_scatter}  shows a scatter plot of such rise times for an ensemble of radio light curves computed over a range of spherically symmetric equivalent mass-loss rates [$\dot{\cal M}(\theta_{\rm obs}) = 10^{-5}$, $10^{-4}$, $10^{-3}$, and $10^{-2} M_{\odot}~{\rm yr^{-1}}$], kinetic energies of the ejecta  [$3\times 10^{50}$, $10^{51}~{\rm erg}$, and $3\times10^{50}$], and microphysical parameters [$(\epsilon_e,\epsilon_B) = (10^{-2},10^{-3})$, $(10^{-3},10^{-4})$, and $(10^{-4},10^{-5})$]. We can see that SNe with spherical CSM  cluster in the region with $\Delta t_{0.01-0.1}/t_{0.1} < 0.3$ and $\Delta t_{0.1-p}/t_{\rm{peak}} < 0.65$, whereas SNe with CSM of other shapes lie outside this region. An exception to this trend is represented by SNe with   $\dot{\cal M}(\theta_{\rm obs})\gtrsim 10^{-2}\,M_{\odot}\,{\rm yr}^{-1}$. In such cases, the shock velocity decreases rapidly, and the impact of free-free absorption is no longer prominent, therefore there is not  a clear distinction of the geometry in Fig.~\ref{fig:LC_scatter} (cf. open symbols). 

We stress that our method holds  as long as free-free absorption dominates over synchrotron self-absorption in the observed band. For $\epsilon_e \gtrsim 10^{-1}$ and $\epsilon_B \gtrsim 10^{-2}$, our method is no longer applicable because synchrotron self-absorption  becomes important (cf.~also Fig.~\ref{fig:radio_eps}).
Moreover, relying on a one-zone model with constant shock velocity, Ref.~\cite{2016MNRAS.460...44P} pointed out that contributions to the radio light curve may also come from secondary electrons generated by proton-proton interactions for $\epsilon_p \gtrsim 10^{-1}$. We neglect this contribution, assuming that  the hadronic contribution  is expected to  decay faster than that of primary electrons for $\epsilon_e \sim \epsilon_p \lesssim 10^{-1}$. 

\subsection{Complementary information  from X-rays}
In principle, the X-ray signal alone is not sufficient to distinguish among the different CSM shapes, because  the X-ray spectrum and its light curve strongly depend  on the spherically symmetric equivalent mass-loss rate and the  microphysical parameter $\eta_{\rm e, th}$ (cf.~Fig.~\ref{fig:Xray_Mdot}). However, once the spherically symmetric equivalent mass-loss rate is constrained along the line of sight from radio observations,  X-ray data can corroborate the radio constraints on the disfavored  CSM configurations. 

For example, if the spherically symmetric equivalent mass-loss rate inferred from radio data along the line of sight is $\dot{\cal M}(\theta_{\rm obs})\sim 10^{-5}\, M_{\odot}\,{\rm yr}^{-1}$, but the RS contribution is not dominant at $t > 10^7\,{\rm s}$ after the explosion, the CSM is likely denser in directions other than the line of sight, as shown in Fig.~\ref{fig:Xray_Mdot}. Furthermore, if the predicted X-ray luminosity based on the SN parameters obtained from the radio fitting cannot reproduce the X-ray data with a reasonable $\eta_{\rm e,th}$, the CSM geometry inferred from radio data should be considered as disfavored.

\subsection{Wind vs.~non-wind circumstellar medium profile: insights from the decaying part of the radio light curve}
Deviations of the radio light curve from the standard spherical  CSM with a wind profile ($s = 2$) can be due to an asymmetric CSM, as explored in this paper, but also to  $s \neq 2$ (see Eq.~\ref{eq:mass_distribution}). Although the rising part of the radio light curves predicted for an asymmetric CSM with $s = 2$ and a spherical CSM with $s \neq 2$ can be similar, their declining phases exhibit different slopes. 

\begin{figure*}
    \centering       
   \includegraphics[width=0.99\columnwidth]{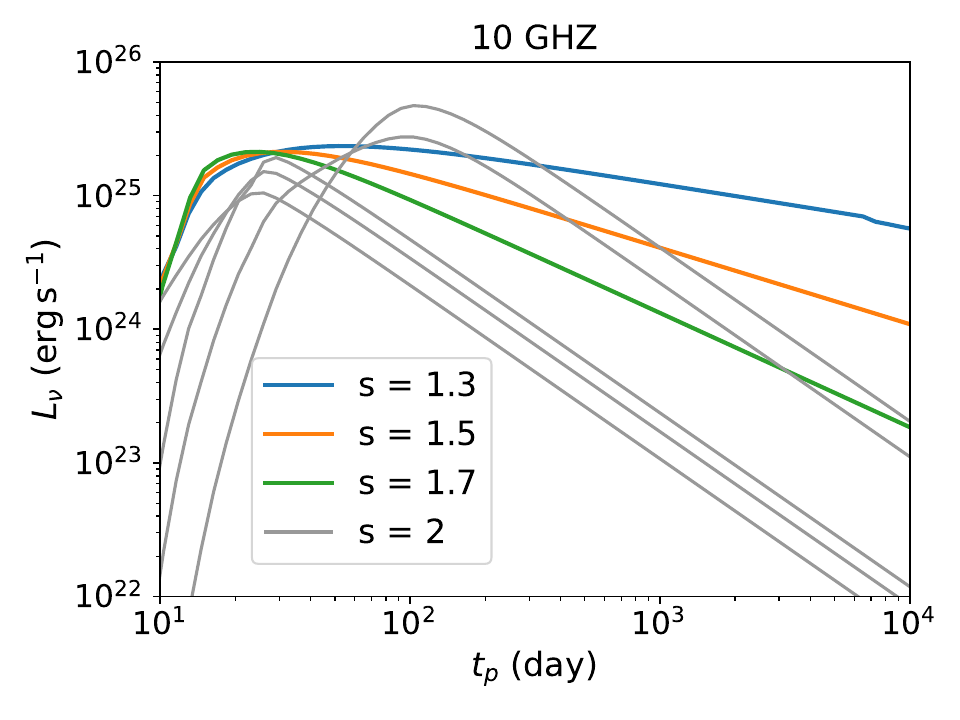}    
    \includegraphics[width=0.99\columnwidth]{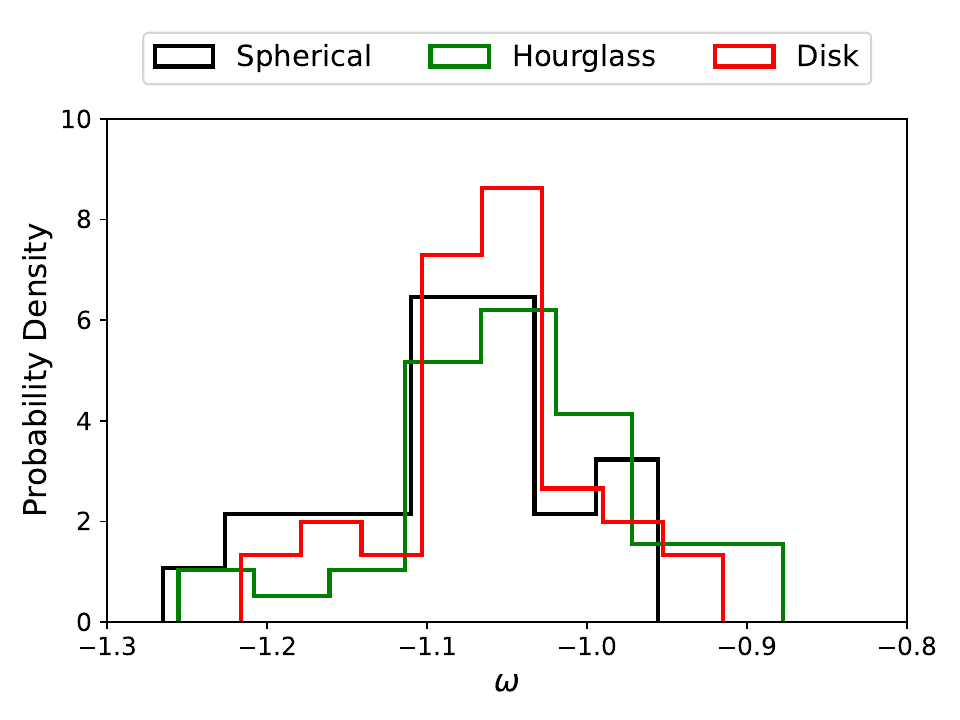} 
    \caption{{\it Left panel}: Radio light curves at $10\,{\rm GHz}$ obtained for  different CSM density profile index: $s=1.3$ in blue, $s=1.5$ in orange, $s=1.3$ in green, and $s=2$ in gray. We  assume a spherical CSM, $\dot{\cal M}(\theta_{\rm obs}) = 10^{-5}\, M_{\odot}\,{\rm yr}^{-1}$, and fix all other parameters as  in Table~\ref{tab:benchmark_value}. The gray lines with $s=2$ account for spherical, hourglass and disk CSM shapes, as well as $\theta_{\rm obs}= 0^\circ$ and $\theta_{\rm obs}= 90^\circ$. {\it Right panel:} Histogram of the distributions of $\omega$, defined as in Eq.~\ref{eq:omega_param}, for $s=2$ and various CSM geometries (spherical in black, hourglass in green, and disk in red). The part of the radio light curve after its peak is very sensitive to $s$ and is essential to discriminate between a non-spherical CSM and a spherical one with $s \neq 2$. }
    \label{fig:decay}
\end{figure*}

Figure~\ref{fig:decay} (left panel) displays examples of light curves computed at $10$~GHz  for  $\dot{\cal M}(\theta_{\rm obs}) = 10^{-5}\,M_{\odot}\,{\rm yr}^{-1}$.  Different values of $s$ are considered,  with all other parameters fixed as in Table~\ref{tab:benchmark_value}.  Comparing the colorful light curves with the gray ones (obtained for $s=2$ and for spherical, disk, and hourglass CSM geometries and our two selected $\theta_{\rm obs}$), we can see that the decay slope of the light curve is more sensitive to  $s$ than to the CSM geometry.

In order to explore the dependence of the radio light curve on $s$ after its peak, we introduce the parameter 
\begin{eqnarray}
    \omega = \frac{{\rm lg}(L_{\rm 0.1,dec}/L_{\rm{peak}})}{{\rm lg}(t_{\rm 0.1,dec}/t_{\rm peak})}\, ,
    \label{eq:omega_param}
\end{eqnarray}
where $L_{0.1, {\rm dec}}/L_{\rm{peak}} = 0.1$ accounts for a decrease of $10\%$ of the luminosity after its peak, and $t_{0.1, {\rm dec}}$ is the time at which $L_{0.1, {\rm dec}}$ is reached. The right panel of Fig.~\ref{fig:decay} shows  that  $\omega$ for SNe with CSM slope $s = 2$ cluster around $\omega \sim -1$, independent of the CSM shape. On the other hand, $\omega$  is smaller  as $s$ decreases and is not degenerate with the CSM geometry.

\section{The circumstellar medium shape of SN 1993j and SN 2023ixf}
\label{sec:applications}
We now apply our findings to two SNe, SN 1993j and SN 2023ixf, that have an extended CSM with a non-trivial structure. 

\subsection{SN 1993j}
Early-time spectroscopy and spectropolarimetry of SN~1993j suggest that the ejecta and the shock interaction region were likely asymmetric~\cite{1994MNRAS.266L..61S,1993ApJ...414L..21T}. Follow-up spectropolarimetric observations also support an asymmetric CSM~\cite{1997PASP..109..489T}. However,  Very Long Baseline Interferometry (VLBI) imaging revealed that the radio shell remained nearly spherical~\cite{1995Natur.373...44M, 2025arXiv250920601F}. A detailed analysis of the optical spectra spanning from the early to late times further suggested that the ejecta were clumpy at early times, while  the CSM interaction became consistent with emission  from a roughly spherical shell after about $433$~days~\cite{2000AJ....120.1499M}. This apparent discrepancy can be reconciled if the observed asymmetry arises  from density inhomogeneities within the shocked shell, rather than from a global asymmetry in the ejecta. Reference~\cite{1996ApJ...461..993F} adopted a spherical CSM model with a density profile shallower than the standard wind one ($s = 1.5$) to reproduce the early radio light curve. Alternatively, such trend in the radio light curve may be mimicked by a clumpy CSM with a volume filling factor.

We reexamine the radio data  provided in Ref.~\cite{1996ApJ...461..993F} and compute  the characteristic rise times ($\Delta t_{0.01-0.1}$ and $\Delta t_{0.1-{\rm peak}}$, cf.~Sec.~\ref{sec:strategy}). We find from Fig.~\ref{fig:LC_scatter} that the radio light curve of SN~1993j can also be explained by an hourglass CSM viewed from $\theta_{\rm obs} = 90^{\circ}$ for a wind density profile ($s = 2$). 

\begin{figure*}
    \centering
    \begin{tabular}{cc}
        \includegraphics[width=0.99\columnwidth]{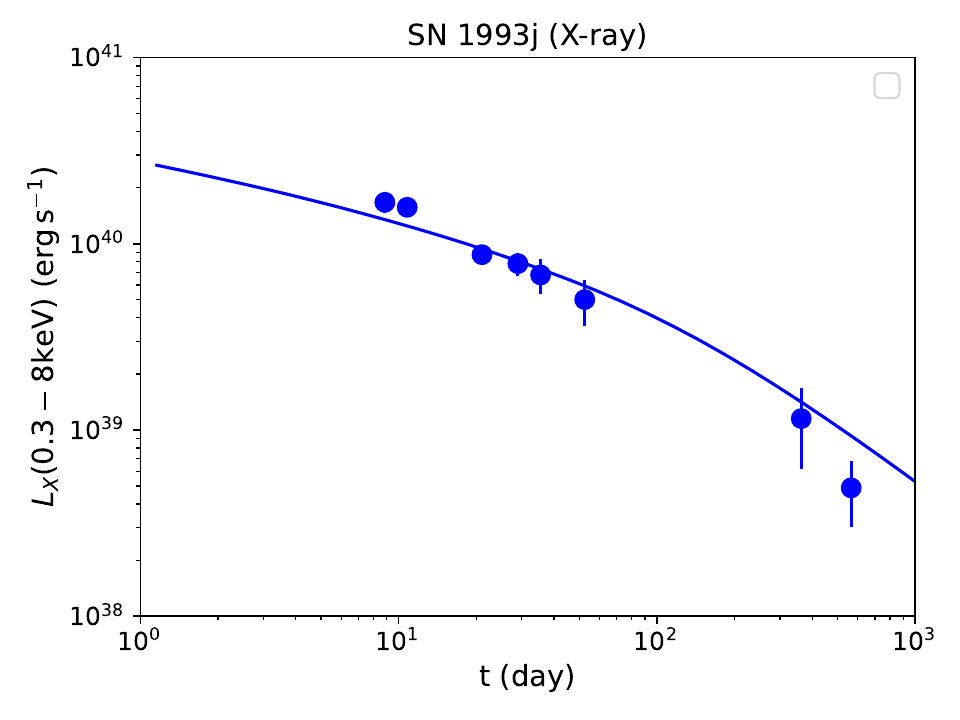}  &   \includegraphics[width=0.99\columnwidth]{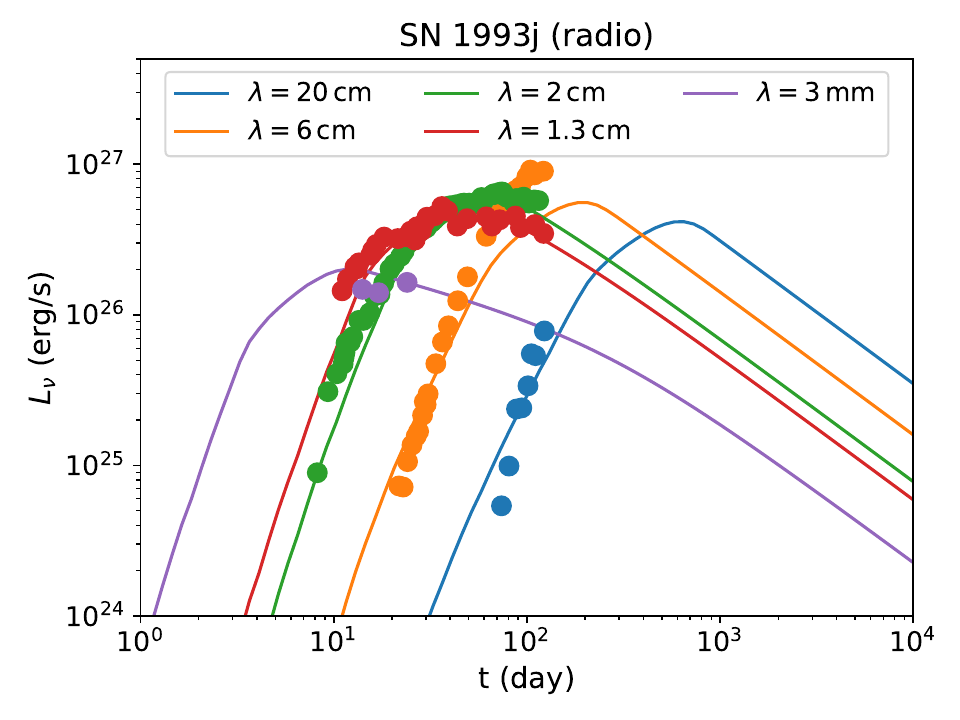}  
    \end{tabular}
    \caption{X-ray in the energy range $0.3$--$8$~keV (on the left) and multi-wavelength radio (on the righ) light curves of SN 1993j for a CSM with hourglass shape and as seen by a distant observer located at $\theta_{\rm obs} = 90^{\circ}$. The SN model parameters that fit observations are provided in Table~\ref{tab:fitted_value}. 
    The radio data at multiple wavelengths have been extracted from Ref.~\cite{1996ApJ...461..993F} and  references therein. The  X-ray data  are from Ref.~\cite{2009ApJ...699..388C}. 
    Our SN model with the hourglass CSM shape is in excellent agreement with both radio and X-ray observations.}
    \label{fig:LC_1993J}
\end{figure*}

Adopting the hourglass CSM profile with $s = 2$ and fixing the other model parameters as in Table~\ref{tab:fitted_value}, the left panel of Fig.~\ref{fig:LC_1993J} provides the multi-wavelength fits of the radio data from Ref.~\cite{1996ApJ...461..993F}. We can see that our model is in excellent agreement with the data. 

Using the SN best-fitting parameters derived from radio  (cf.~Table \ref{tab:fitted_value}), we model the X-ray emission and compare the latter with the X-ray data from Ref.~\cite{2009ApJ...699..388C}. Our findings are shown in the right panel of Fig.~\ref{fig:LC_1993J}, and our model is in  excellent agreement with the radio data. 
The integrated X-ray flux in the $0.3$--$8\, {\rm keV}$ band is dominated by the RS contribution.

We note that, with these fitted parameters, the interaction shell can be quasi-spherical, as the ejecta expand isotropically and the CSM density is too low to affect the expansion; angular variations in the CSM density only have an impact on the emission, but not on the shell geometry. However, VLBI observations have shown that the expansion of the radio ring decelerated after 
about $5$~years~\citep{2000Sci...287..112B}, suggesting that the swept CSM mass is large enough to affect the shock dynamics. Hence, the approximately spherical geometry would suggest a nearly spherical CSM. On the other hand, the brightness distribution of the radio ring observed by VLBI deviates from spherical symmetry~\citep{2000Sci...287..112B, 2001ApJ...557..770B, 2003ApJ...597..374B}:  this suggests that the CSM or the SN ejecta could be asymmetric. This  apparent conflict requires further investigation; it is possible that the CSM is moderately asymmetric and has a non-wind radial density profile ($s < 2$)--both effects would jointly shape the lightcurves.

\subsection{SN 2023ixf}
High-resolution spectra of SN~2023ixf during the first week after the explosion reveal blueshifted emission lines, indicating a pre-shocked CSM with an expansion velocity higher than the typical supergiant wind velocity~\cite{2023ApJ...956...46S}. The detection of a high-velocity broad absorption component in the blue wing of the Balmer lines favors an asymmetric shock geometry~\citep{2024ApJ...975..132S}. Spectropolarimetric observations also favor such a scenario~\cite{2023ApJ...955L..37V, 2025ApJ...982L..32S, 2026ApJ..1000...18V}. The shape of the radio light curve of SN~2023ixf deviates from that expected for a spherically symmetric CSM with a wind density profile ($s = 2$). Assuming a spherical CSM, Ref.~\citep{2025ApJ...985...51N} instead adopted $s = 1.3$. 

Assuming $s=2$, the characteristic rise times (see Sec.~\ref{sec:strategy}) for the radio data are shown in Fig.~\ref{fig:LC_scatter}. Such characteristic times indicate that an asymmetric CSM structure is preferred, although the preferred CSM geometry and viewing angle cannot be uniquely determined.

\begin{figure*}
    \centering
    \begin{tabular}{cc}        \includegraphics[width=0.99\columnwidth]{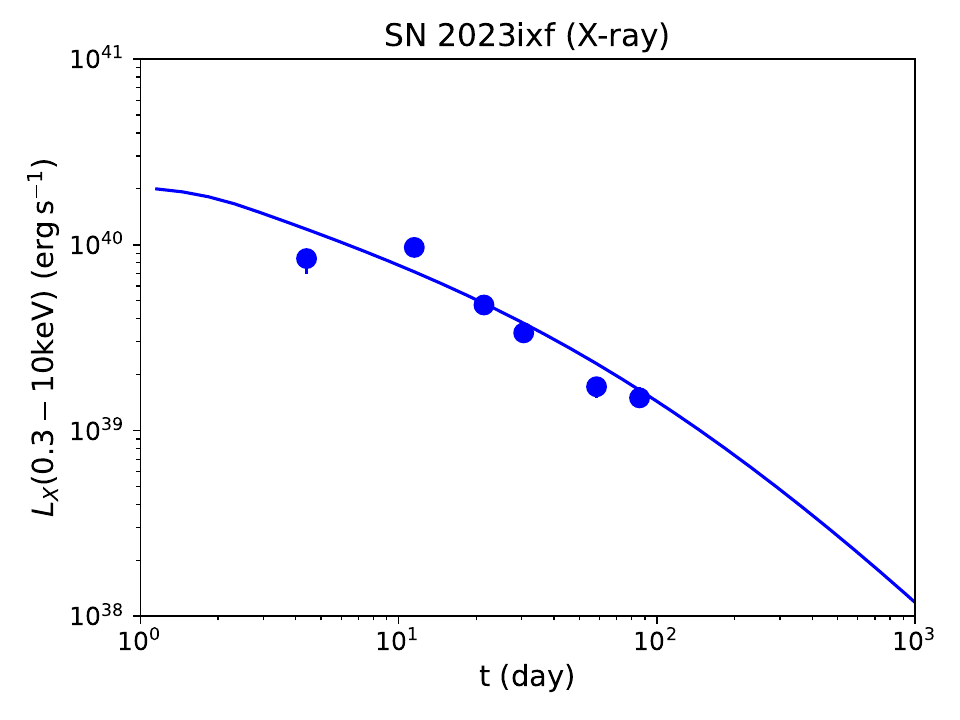}  &   \includegraphics[width=0.99\columnwidth]{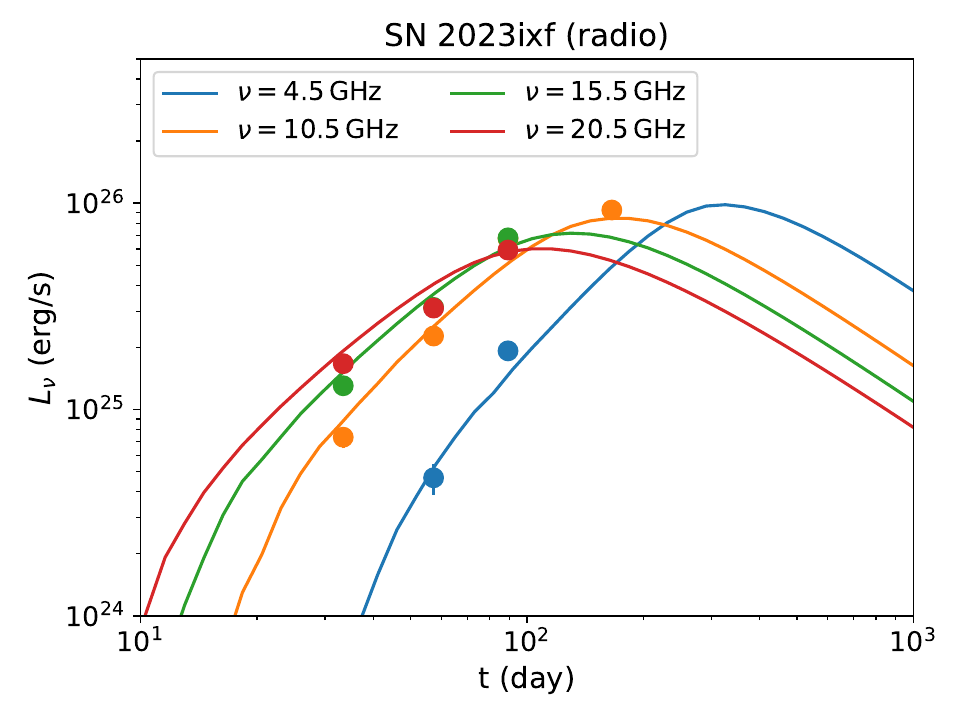}  
    \end{tabular}
    \caption{Same as  Fig.~\ref{fig:LC_1993J}, but for SN 2023ixf. We assume a CSM with hourglass shape and an observer located in the equatorial plane with $\theta_{\rm obs} = 90^{\circ}$.  The SN model parameters that fit observations are provided in Table~\ref{tab:fitted_value}. Note that the RS contribution in negligible in this case, and the  FS contribution is identical to the total X-ray luminosity.
    The radio and X-ray data are extracted from Ref.~\cite{2025ApJ...985...51N} and display very good agreement with our SN model. Our SN model with hourglass CSM is in excellent agreement with X-ray and radio observations. 
    }
    \label{fig:LC_2023ixf}
\end{figure*}

Motivated by the findings of Ref.~\citep{2025ApJ...982L..32S}, we explore the possibility that this SN has a CSM with hourglass shape and a wind density profile ($s = 2$). We assume that SN 2023ixf is observed from  $\theta_{\rm obs}=90^{\circ}$. Figure~\ref{fig:LC_2023ixf} shows our multi-wavelength light curves in radio and  X-rays in the $0.3$--$10\,{\rm keV}$ band obtained for the SN model parameters in Table~\ref{tab:fitted_value}. Our model is in excellent agreement with the observational data from Ref.~\cite{2025ApJ...985...51N}. Moreover, also in this case, we find that the X-ray light curve is dominated by the FS contribution.   

\begin{table}[ht]
\centering
\caption{Model parameters adopted to fit the light curves in X-rays and radio of  SN 1993j and SN 2023ixf; cf.~Figs~\ref{fig:LC_1993J} and \ref{fig:LC_2023ixf}.}
\renewcommand{\arraystretch}{1.3}
\begin{threeparttable}
\resizebox{\columnwidth}{!}{%
\begin{tabular}{lcc}
\hline
Physical Quantity & SN 1993j &  SN 2023ixf \\
\hline
\multicolumn{3}{l}{\textit{Ejecta parameters}} \\
\hline
Kinetic energy ($E_{\rm ej}$) & $5.0\times 10^{51}~{\rm erg}$ & $2.0 \times 10^{51}~{\rm erg}$ \\
Mass ($M_{\rm ej}$) & $10\, M_{\odot}$ & $10\, M_{\odot}$ \\
Outer  density index ($n$) & $12$ & $12$ \\
Inner  density index ($\delta$) & $0.5$ & $0.5$ \\
\hline
\multicolumn{3}{l}{\textit{CSM parameters}} \\
\hline
Mass-loss rate ($\dot{\cal M}(\theta_{\rm obs})$) & $5.0\times 10^{-6} M_{\odot}/{\rm yr}$ & $1.5\times 10^{-5} M_{\odot}/{\rm yr}$ \\
Wind velocity ($v_{\rm w}$) & $25~{\rm km~s^{-1}}$ & $25~{\rm km~s^{-1}}$  \\
Radial density index ($s$) & $2$ & $2$ \\
Hourglass  parameter ($A$) & $10$ & $20$ \\
Hourglass  parameter ($\beta$) & $2.2$ & $1.2$ \\
\hline
\multicolumn{3}{l}{\textit{Radiation parameters}} \\
\hline
Electron heating efficiency ($\eta_{\rm e, th}$) & $0.3$ & $0.007$ \\
Electron energy fraction ($\epsilon_e$) & $5.0\times10^{-2}$ & $3.5\times10^{-2}$ \\
Magnetic energy fraction ($\epsilon_B$) & $10^{-4}$ & $10^{-4}$ \\
Electron power-law index ($p$) & $2.3$ & $3$ \\
\hline
\end{tabular}%
}
\end{threeparttable}
\label{tab:fitted_value}
\end{table}

\section{Discussion and conclusions} 
\label{sec:conclusion}
Interacting SNe are surrounded by a dense CSM originating from the wind ejected throughout the star lifetime as well as impulsive mass-loss episodes. A growing set of data provides  evidence that the CSM can be characterized by large asymmetries across the emission directions. Understanding the properties and the shape of the CSM is crucial to interpret multi-messenger observations and gather insight on the mass-loss history of the SN progenitor.

In this paper, we model the X-ray and radio spectra and light curves from interacting SNe, assuming that the CSM has a spherical, hourglass, or disk shape. 
We find that the spectral shape, its temporal evolution, and the shape of the light curves are strongly affected by the CSM geometry,  even when the same density profile along the observer’s line of sight is adopted.

As for the X-ray emission, we account for contributions from  the FS and the RS. As the radiation produced at the RS propagates through the cool dense shell between the RS and  the contact discontinuity, it  can undergo bound–free absorption in the ultraviolet and soft X-ray bands. Nevertheless, the RS component dominates the  X-ray emission when the CSM density is relatively low.

The radio emission generated by the FS can experience  free–free absorption while propagating through the unshocked ionized medium. The efficiency of free-free absorption depends on the direction of emission for an asymmetric CSM. Therefore  the observer sees the superposition of direction-dependent features due to free-free absorption, responsible for a deviation of the  synchrotron spectrum  from its standard shape. 
In particular, we find a flattening of the synchrotron spectrum near the peak frequency. The slope of the rising phase in the radio light curve varies with the CSM geometry. When the  CSM density along the line of sight is larger than the density along other directions, the radio light curve has a shallower  rise with respect to  the case of spherical CSM. If the CSM density along the line of sight  is lower than the density along other directions, the   slope of the rising part of the light curve is comparable to the one of the spherical CSM model.

Building on the dependence of the  radio signal on the CSM shape, we propose a strategy to  infer the CSM geometry from observations. To this purpose, we introduce two characteristic timescales: the timescale for the luminosity to vary by $10\%$ with respect to its peak  ($\Delta t_{0.1-{\rm peak}}$) and the timescale for the luminosity to increase from $1\%$ to $10\%$ of the peak luminosity ($\Delta t_{0.01-0.1}$). A  plot of these two  characteristic timescales reveals that, for an ensemble of SNe with different properties, these characteristic times tend to be similar for SNe with the same CSM shape. Moreover, the   decay phases of the radio light curve are useful to discriminate  between  wind vs. non-wind CSM profiles. Once the CSM shape is inferred from radio observations, the X-ray signal can be used for  an independent cross check. 

We have tested our method relying on the X-ray and radio data sets from two of the most studied SNe with an asymmetric CSM: SN 1993j and SN 2023ixf. It was suggested in the literature that  a spherical CSM with a non-wind density profile ($s \neq 2$) could be in agreement with observations~\citep{1996ApJ...461..993F,2025ApJ...985...51N}. However, we find that an asymmetric CSM with a wind density profile ($s = 2$) can also reproduce the multi-wavelength light curves if the CSM has an  hourglass shape and both SNe were observed  from the equatorial plane ($\theta_{\rm obs} = 90^{\circ}$).  The X-ray observations of both SNe are in agreement with our model based on the radio data.

We stress that our results are based on a  model representative of Type II SNe. Although we have scanned a wide range of the parameter space, our conclusions may change if SNe have lower ejecta mass or larger spherically symmetric equivalent mass loss.  Our strategy to discriminate among different CSM shapes is  valid when the CSM shape deviates significantly from the spherical geometry, specifically for $\theta_{\rm CSM} < 60^{\circ}$ and $m > 1$ for the disk shape, as well as for $A > 3$ and $\beta > 0.5$ for the hourglass shape. Moreover, our method holds as long as synchrotron self-absorption is negligible (i.e., for $\epsilon_e \lesssim 10^{-1}$ and $\epsilon_B \lesssim 10^{-2}$). 
We stress that, for higher particle acceleration efficiency and when hadronic processes are taken into account, the signatures of the CSM geometry around the peak of the radio light curve  may be further modified by synchrotron self-absorption and secondary electrons. 

One caveat of our X-ray emission modeling is that we adopt a simplified treatment for inverse Compton scattering by non-relativistic and relativistic electrons. Although we account for the  shock power to compute inverse Compton cooling and implement this cooling term in our model, we do not account for the resulting X-ray radiation. In fact, reproducing the Compton-scattered X-ray emission requires the seed photon spectrum, which in turn demands for self-consistent calculations across the optical, ultraviolet, and X-ray bands. Comptonization is investigated in Refs.~\cite{2022ApJ...928..122M,2025ApJ...993...46W}--the latter presents  numerical simulations of radiation-mediated shocks. Both studies conclude that Comptonization by thermal electrons becomes non-negligible for shock velocities exceeding $\sim 10^9\,{\rm cm\,s^{-1}}$, whereas bremsstrahlung dominates at lower velocities. Our model parameter set falls within the bremsstrahlung-dominated regime, therefore our results are in good agreement with the ones in Ref.~\cite{2025ApJ...993...46W}. The inverse Compton emission from relativistic electrons may be stronger, as pointed out in Ref.~\cite{2025ApJ...985...51N}, yet its strength depends on the particle acceleration efficiency and warrants further investigation. Beyond X-ray emission, Ref.~\cite{2022ApJ...928..122M}  accounts for bound-free absorption arising from neutral CSM;  we assume that the CSM is fully ionized by photons generated in the shock, and thus omit bound-free absorption from the unshocked CSM in our calculations.  On the other hand, we compute the X-ray emission originating from the RS, a component neglected in Ref.~\cite{2022ApJ...928..122M}. The RS emission can dominate the soft X-ray band for SNe with relatively low mass-loss rates, even when strong bound-free absorption occurs in the cool dense shell.
Our calculations show that the RS upstream remains optically thick after the FS breakout, making the RS radiation mediated. Radiation smooths the RS velocity jump, weakening particle acceleration~\cite{2013PhRvL.111l1102M,Ai:2025qst,Budnik:2010ru}; hence we only include the FS synchrotron emission.

Our strategy can be applied to a specific SN event; however  differences in the  treatment of relevant physical processes can alter our results. For example, 
Ref.~\cite{1996ApJ...461..993F} claims that the radial density profile of the CSM of SN 1993J has an index of $s = 1.5$-$1.7$, whereas Ref.~\cite{1998ApJ...509..861F} accounts for synchrotron self-absorption and finds that $s = 2$ explains the observed radio light curves, with synchrotron self-absorption playing a significant role. A key point is whether free-free absorption or synchrotron self-absorption dominate at relatively long wavelengths, sensitive to the temperature of the unshocked CSM. In our model, we fix $T_{\rm CSM} = 10^5\,{\rm K}$, in which case free-free absorption is the dominant process. Our conclusions may be affected if one were to consider variations in the CSM temperature. In addition, the fitted electron heating efficiency ($\eta_{e,{\rm th}}$) is different for SN 1993j and SN 2023ixf, while the microphysical parameters are assumed to be universal. This is because, in this work, the initial kinetic energy and  ejecta mass are fixed, while the downstream electron temperature is determined by both the shock velocity and the electron heating efficiency. Such  efficiency should be adjusted to match the observed X-ray radiation temperature. If variations in the shock velocity are allowed, the electron heating efficiencies in different SNe may appear to be more similar.

We leave to future work a detailed treatment of the composition and ionization of the CSM and SN ejecta to obtain more realistic results, see e.g.~Ref.~\citep{2006A&A...449..171N} for preliminary work in this direction. More complex CSM structures could also be considered; for example, the radial density profile index $s$ could be allowed to vary with both radius and angle. Even if the non-spherical CSM model is able to explain the multi-wavelength observations, further investigation on the  CSM properties is needed~\cite{2026arXiv260620836W}.

Our findings highlight the crucial role played by radio and X-ray observations within a multi-messenger framework. The strategy outlined in this work is crucial to gain insight on the growing number of puzzling observations of the final stages of  the life of massive stars.

\begin{acknowledgments}
We are especially grateful to Raffaella Margutti for insightful discussions and feedback on the manuscript. This project has received support from the Villum Foundation (Project No.~13164).
\end{acknowledgments}

\bibliography{bibliography}

\end{document}